\documentclass[runningheads,a4paper]{llncs}

\usepackage[T1]{fontenc}
\usepackage{stmaryrd}
\usepackage{algorithm}
\usepackage{algorithmic}
\usepackage{graphicx}
\usepackage{floatflt}
\usepackage{wrapfig}
\usepackage{pstricks}
\usepackage{rotating}
\usepackage{subfigure}
\usepackage{amsmath}
\usepackage{amssymb}
\usepackage{theorem}
\usepackage[active]{srcltx} 
\usepackage{amsmath,amssymb}
\usepackage{booktabs}
\usepackage{xcolor}
\usepackage{hyperref}
\usepackage{listings}
\usepackage{enumerate}
\usepackage[justification=RaggedRight, singlelinecheck=false]{caption}
\usepackage{marvosym}

\newcommand{\QED}{\hspace*{\fill}$\Box$}

\newcommand{\T}{{\cal T}}
\newcommand{\K}{{\cal K}}
\newcommand{\A}{{\cal A}}
\newcommand{\B}{{\cal B}}
\newcommand{\C}{{\cal C}}
\newcommand{\D}{{\cal D}}

\newcommand{\union}{\cup}
\newcommand{\Pred}{{\sf Pred}}
\newcommand{\ignore}[1]{}
\newcommand{\alg}{\cal} 
\newcommand{\pK}{\Psi_{\K}}

\newenvironment{enumerate-} 
{\begin{enumerate}
    
   \setlength{\parskip}{-1ex}              
   \setlength{\itemsep}{1.5ex}             
}
{
 \end{enumerate}
}

\newtheorem{defi}{Definition}
\newtheorem{thm}{Theorem}
\newtheorem{lem}[thm]{Lemma}

\newtheorem{cor}[thm]{Corollary}
\newtheorem{prop}[thm]{Proposition}
\newtheorem{rem}{Remark}
\newtheorem{ex}{Example}

\begin{document}
\pagestyle{headings}
\title{On Symbol Elimination and Uniform Interpolation in Theory Extensions}

 \titlerunning{  }
\author{Viorica
  Sofronie-Stokkermans\textsuperscript{(\Letter)}\orcidID{0000-0002-8486-9955}}
\institute{University of Koblenz, Koblenz, Germany}
\maketitle

\begin{abstract}
We define a notion of general uniform interpolant, generalizing the
notions of cover and of uniform interpolant and identify situations in
which symbol elimination can be used for computing general uniform 
interpolants. We investigate the limitations of the method we propose, 
and identify theory extensions for which the computation of general 
uniform interpolants can be reduced to symbol elimination followed 
by the computation of uniform quantifier-free interpolants in
extensions with uninterpreted
function symbols of theories
allowing uniform quantifier-free interpolation.
\end{abstract}

\section{Introduction}

In this paper we study the links between symbol elimination in
local theory extensions and 
possibilities of computing uniform interpolants in such theories. 

We studied possibilities of symbol elimination in local theory 
extensions in \cite{sofronie-ijcar10,sofronie-cade13,peuter-sofronie19},  in
relationship with the verification of safety properties in 
parametric systems: 
symbol elimination was used to 
infer (possibly universally quantified) constraints
on parameters under which it was guaranteed 
that certain properties are met, or for invariant strengthening. In \cite{sofronie-ijcar16,sofronie-lmcs18} we 
continued this study and analyzed possible applications to interpolation 
in local theory extensions.

Interpolation is widely used in program verification
(cf.\
\cite{McMillanCAV03,McMillanProver04,McMillanSurvey05,McMillanRelationApproximation,Kapur-et-all-06}
to mention some of the early papers). 
Intuitively, interpolants can be used for describing separations 
between the sets of ``good'' and ``bad'' states; they can 
help to discover relevant predicates in predicate abstraction with 
refinement and for over-approximation in model checking. 
It often is desirable to obtain ground interpolants of 
ground formulae. 
When defining ground interpolation, sometimes a difference is
made between interpreted and uninterpreted function symbols or
constants. In  \cite{BruttomessoGR14}, a notion of {\em general ground
  interpolation} was defined. 
A theory ${\cal T}$ with signature $\Sigma$ is said to have the 
{\em general ground interpolation property} if for every signature
$\Sigma'$ disjoint from $\Sigma$ 
and every pair of ground $\Sigma \cup \Sigma'$-formulae $A$ and  
$B$,  if $A \wedge B \models_{{\T} \cup {\sf UIF}_{\Sigma'}} \perp$ then 
there exists a ground formula $I$ such that: 
(i) all constants and all predicate and function symbols from $\Sigma'$ 
occurring in $I$ are symbols shared by $A$ and $B$ (no restriction is imposed on 
the symbols in $\Sigma$),  and 
(ii) $A \models_{{\mathcal T} \cup {\sf UIF}_{\Sigma'}} I \text{ and } 
B  \wedge I \models_{{\mathcal T} \cup {\sf UIF}_{\Sigma'}}  \perp.$
The symbols in $\Sigma$ are regarded as interpreted, those in $\Sigma'$ as
uninterpreted. 
In \cite{Sofronie-ijcar-06,sofronie-lmcs08,sofronie-ijcar16,sofronie-lmcs18,peuter-sofronie-cade-2023} 
we studied theory extensions $\T_0 \subseteq \T_0 \cup \K$
for which the notion of sharing had to be extended, because of
uninterpreted function symbols occurring 
together in clauses in $\K$. We solved this by introducing 
closure operators $\Theta$ on signatures and imposing that the 
interpolant contains terms in $\Theta(\Sigma_s \cup C_s)$, where 
$\Sigma_s \cup C_s$ are the function, predicate symbols and constants
shared by $A$ and $B$.

Sometimes one needs to find the {\em strongest} interpolant of two
formulae w.r.t.\ a theory ${\cal T}$, 
or a {\em (general) uniform interpolant} for a ground formula $A$ w.r.t.\
a given subset $\Sigma_s$ of function symbols and a subset $C_s$ of
the constants occurring in $A$ for a theory ${\cal T}$. This is a formula $I$ containing 
arbitrary interpreted symbols and only uninterpreted 
symbols in $\Sigma_s \cup C_s$ (or a suitably defined closure
$\Theta(\Sigma_s \cup C_s)$ depending on the theory ${\cal T}$), such
that $A \models_{\mathcal T} I$ and for all ground formulae $B$ 
sharing with $A$ only interpreted symbols and symbols in $\Sigma_s
\cup C_s$ (resp.\ $\Theta(\Sigma_s \cup C_s)$) with 
$A \models_{\mathcal T} B$, we have $I \models_{\mathcal T} B$. 
Uniform interpolants were studied in the context of non-classical
logics and description logics and in verification. In description logics 
they are studied in relationship with ``forgetting'' (projecting
to a subsignature) (cf.\ e.g.\ \cite{konev-wolter-09,KonevWW09,lutz-wolter-ijcai11}). 
In verification, uniform interpolants offer an optimal
approximation of states in a subsignature; in fact, a strongly
related notion -- that of 
cover -- was proposed by Gulwani and Musuvathi in
\cite{Gulwani-Musuvathi}. The computation of uniform interpolants
allows a more precise approximation or even a precise computation of
the reachable states. 
The problem of uniform interpolation for the theory of uninterpreted 
function symbols was studied in
\cite{ghilardi-superposition,ghilardi-dag}, 
where several methods were proposed: a method using a version of constraint
superposition \cite{ghilardi-superposition}, DAG-methods and tableau methods \cite{ghilardi-dag}. 
In \cite{ghilardi-combinations} possibilities of uniform interpolation in combinations of
theories are investigated. 

However, the results in \cite{Gulwani-Musuvathi,ghilardi-superposition,ghilardi-dag,ghilardi-combinations} are
devised for the case in which all function symbols are seen as 
interpreted and only constants need to be eliminated: The computed 
interpolant can contain arbitrary function symbols, but only constants 
in a specified set $C_s$. 
In this paper we extend these results and investigate the possibility
of generating general uniform interpolants which contain uninterpreted
symbols in specified subsets $\Sigma_s \cup C_s$ (resp.\ in 
$\Theta(\Sigma_s \cup C_s)$ for a suitable closure operator), 
identify classes of theory extensions where {\em general uniform
  interpolation} is possible  and also investigate the limitations
of the method we propose. The main contributions of the paper 
can be summarized as follows: 
\begin{itemize}
\item We define a notion of general uniform interpolant generalizing
  the notions of cover \cite{Gulwani-Musuvathi} and of uniform
  interpolant \cite{ghilardi-superposition} and prove a 
  semantical characterization of general uniform
  interpolants, extending the one in 
  \cite{ghilardi-superposition}. 
\item We propose a method for eliminating function symbols with arity
  $\geq$ 1 based on an adaptation of a method for symbol elimination 
we proposed in \cite{sofronie-lmcs18}. 
\item We prove that certain extensions of theories with uninterpreted
  function symbols have general uniform interpolation, and 
identify theory extensions for which 
the computation of {\em general uniform interpolants} can be reduced to 
the computation of {\em uniform interpolants} 
with methods from \cite{ghilardi-superposition,ghilardi-dag,ghilardi-combinations}.

\item We propose a method for computing explicit definitions for
  functions implicitly definable w.r.t.\ the signature of the base
  theory + uninterpreted function symbols in local theory extensions, and analyze possibilities of
  computing general uniform interpolants in such
  extensions. 
\end{itemize}
{\em The paper is structured as follows:} In Section~\ref{prelim} the
main definitions and results on logical theories and local theory
extensions which are used in the paper are introduced. 
In Section~\ref{unif-int} we recall existing results on $\T$-uniform
quantifier-free interpolation. In Section~\ref{sect:gen-unif-int} we define a notion of 
general uniform interpolant w.r.t.\ a theory ${\cal T}$. 
In Section~\ref{symbol-elimination} we present a method for 
symbol elimination and investigate conditions under which it 
can be used for computing general uniform interpolants. 
In Section~\ref{definitions} we analyze extensions with 
implicitly definable functions.
Section~\ref{conclusions} contains the conclusions and some plans of
future work. 

\smallskip
\noindent This paper is the extended version of \cite{sofronie-cade-2025}: 
it provides details of proofs and additional examples.

\section*{Table of Contents}
\contentsline {section}{\numberline {1}Introduction}{1}{section.1.1}%
\contentsline {section}{\numberline {2}Preliminaries}{3}{section.1.2}%
\contentsline {subsection}{\numberline {2.1}Logical Theories}{3}{subsection.1.2.1}%
\contentsline {subsection}{\numberline {2.2}Local Theory Extensions}{5}{subsection.1.2.2}%
\contentsline {section}{\numberline {3}Uniform interpolation}{7}{section.1.3}%
\contentsline {section}{\numberline {4}General uniform interpolation}{9}{section.1.4}%
\contentsline {section}{\numberline {5}General uniform interpolation in theory extensions}{10}{section.1.5}%
\contentsline {subsection}{\numberline {5.1}(I) Eliminating function symbols in local theory extensions}{11}{subsection.1.5.1}%
\contentsline {subsection}{\numberline {5.2}(II) Eliminating constants occurring below function symbols}{15}{subsection.1.5.2}%
\contentsline {subsubsection}{\numberline {5.2.1}Extensions with uninterpreted function symbols.}{16}{subsubsection.1.5.2.1}%
\contentsline {subsubsection}{\numberline {5.2.2}Uniform interpolation with $\Theta $-sharing.}{17}{subsubsection.1.5.2.2}%
\contentsline {section}{\numberline {6}Definable functions}{21}{section.1.6}%
\contentsline {section}{\numberline {7}Conclusions}{27}{section.1.7}%
\contentsline {section}{\numberline {A}Local theory extensions}{31}{section.A.1}%

\section{Preliminaries}

\label{prelim}

We assume known standard definitions from first-order logic  
such as $\Pi$-structures, models, homomorphisms, logical entailment,  
satisfiability, unsatisfiability.  We here introduce the main notions 
needed in the paper:  logical theories and properties of logical
theories and local theory extensions. 

\medskip
\noindent {\em Notation:} In what follows we will denote with (indexed
version of) 
${\overline x}$ (resp. ${\overline c}$) sequences of variables $x_1, \dots,
x_n$ (resp.\ sequences of constants $c_1, \dots, c_n$).

\subsection{Logical Theories} 
We consider many-sorted signatures of the form $\Pi = (S, \Sigma, {\sf Pred})$, 
where $S$ is a set of sorts, $\Sigma$ is a family of function symbols and ${\sf Pred}$
a family of predicate symbols such that for every $f \in \Sigma$ and
$P \in {\sf Pred}$ their 
arities $a(f) = s_1 \dots s_n \rightarrow s$ resp.\ $a(P) =
s_1,...,s_n$ (where $s_1,\dots,s_n, s \in S$) are specified. 
In this paper, a theory $\T$ is described by a set of closed formulae 
(the axioms of the theory). 
If we do not want to emphasize the set of sorts, we write $\Pi =
(\Sigma, {\sf Pred})$. 
We denote
``falsum'' with $\perp$. If $F$ and $G$ are formulae we 
write $F \models G$ (resp. $F \models_{\cal T} G$) 
to express the fact that every model of $F$ 
(resp. every model of $F$ which is also a model of 
$\T$) is a model of $G$. 
The definitions can be extended in a natural way to the case when 
$F$ is a set of formulae; in this case, $F \models_{\cal T} G$ if and only if 
${\cal T} \cup F \models G$. 
$F \models \perp$ means that $F$ is
unsatisfiable; $F \models_{\T} \perp$ means that there is no model of
$\T$ which is also a model of $F$. If there is a model of $\T$ which is also a
model of $F$ we say
that $F$ is $\T$-consistent.
If $C$ is a fixed countable set of fresh constants, we denote by 
$\Pi^C$ the extension of $\Pi$ with constants in $C$.

For $\Pi$-structures ${\cal A}$ and ${\cal B}$, 
$\varphi : {\cal A} \rightarrow {\cal B}$ is an embedding if and only if 
it is an injective homomorphism and has the property that 
for every $P \in {\sf Pred}$ with arity $n$ and all 
$(a_1, \dots, a_n) \in A^n$, $(a_1, \dots, a_n) \in P_\A$ iff 
$(\varphi(a_1), \dots, \varphi(a_n)) \in P_{\cal B}$. 
In particular, an embedding preserves the truth of all literals.
An elementary embedding between two $\Pi$-structures is an 
embedding that preserves the truth of
all first-order formulae over $\Pi$. 
Two $\Pi$-structures are elementarily equivalent if they satisfy 
the same first-order formulae over $\Pi$.

Let ${\cal A} = (A, \{ f_{\A} \}_{f \in \Sigma}, \{ P_\A \}_{P \in {\sf
    Pred}})$ be a $\Pi$-structure. In what follows, we will 
denote the universe $A$ of the structure ${\cal A}$ by $|{\cal A}|$. 
If $\Pi_1 \subseteq \Pi$, we will denote by ${\cal A}_{|\Pi_1}$ the
reduct of ${\cal A}$ to $\Pi_1$, and if $C$ is an additional set of 
constants we will denote by ${\cal A}^C$ the extension of ${\cal A}$
with new constants in $C$ (i.e.\ a $\Pi^C$-structure with ${\cal A}_{|\Pi}
= {\cal A}$ and with interpretations for the additional constants in
$C$).
Let $\A^A$ be the extension of $\A$ where we have an additional 
constant for each element of $A$ (which we here denote with the
 same symbol) with the natural expanded interpretation
mapping the constant $a$ to the element $a$ of $|\A|$. 
The
\emph{diagram} $\Delta(\A)$ of $\A$ is the set of all
literals true in $\A^A$
(it is a set of  sentences over the
signature $\Pi^A$ obtained by expanding 
$\Pi$ with a fresh constant $a$ 
for every element $a$ from $|\A|$).
Note that if ${\cal A}$ is a $\Pi$-structure and $\T$ a theory and 
$\Delta({\cal A})$  is $\T$-consistent then there exists a 
$\Pi$-structure ${\cal B}$ which is a model of $\T$
and into which ${\cal A}$ embeds. 

\smallskip
\noindent {\bf Quantifier elimination, model completeness, model
  completion.} A theory $\T$ over signature  
$\Pi$ {\em allows quantifier elimination} if for every formula $\phi$ over  
$\Pi$ there exists a quantifier-free formula $\phi^*$ over  
$\Pi$ which is equivalent to $\phi$ modulo $\T$. 
A \emph{model complete} theory has the property that
all embeddings between its models are elementary.
Every theory which allows 
quantifier elimination (QE) is model complete (cf.\
\cite{hodges}, Theorem 7.3.1). Thus, since 
linear rational arithmetic and linear real arithmetic $LI({\mathbb Q})$, 
$LI({\mathbb R})$, and the theory $\mathbb{R}$ of real closed fields,
as well as the theory 
of algebraically closed fields 
allow quantifier elimination, they are model complete. 
A theory $\T^*$ is called a \emph{model companion} 
of $\T$ if (i) $\T$ and $\T^*$ are co-theories 
(i.e.\ every model of $\T$ can be extended to a model of 
$\T^*$ and vice versa), (ii) $\T^*$ is model complete. 
$\T^*$ is  a \emph{model completion} of $\T$
if it is a model companion of $\T$ with the additional 
property 
(iii) for every model ${\cal A}$ of $\T$, 
 $\T^* \union \Delta({\cal A})$ is a complete theory (where $\Delta({\cal A})$ is the diagram of $\A$).

\smallskip
\noindent {\bf Stable infinity.} A theory ${\cal T}$ is stably infinite iff for every ground
formula $\phi$, if $\phi$ is $\T$-consistent then there exists also an infinite model
${\cal B}$ of ${\cal T}$ with ${\cal B} \models \phi$.

\smallskip
\noindent {\bf Convexity.}
A theory ${\cal T}$ with signature $\Pi = (\Sigma, {\sf Pred})$ is 
{\em convex} if for all conjunctions $\Gamma$ of ground 
$\Pi^C$-atoms (with additional constants in a set $C$), 
and $\Pi^C$-terms $s_i, t_i$, $i \in \{ 1, \dots, n \}$, 
if $\Gamma \models_{\T} \bigvee_{i = 1}^m s_i \approx t_i$ 
there exists $i_0 \in  \{ 1, \dots, m \}$ such that 
$\Gamma \models_{\T} s_{i_0} \approx t_{i_0}$.

\medskip
\noindent {\bf Equality interpolation.}
We say that a convex theory $\T$ has the equality interpolation property if 
for every conjunction of ground $\Pi^C$-literals 
$A(\overline{c}, \overline{a_1}, a)$ and 
$B(\overline{c}, \overline{b_1}, b)$,  
if $A \wedge B \models_{\cal T} a \approx b$
then there exists a term $t(\overline{c})$ containing only the shared
constants $\overline{c}$ such that $A \wedge B \models_{\cal T} a \approx
t(\overline{c}) ~\wedge~ t(\overline{c}) \approx b$. 
A theory allowing quantifier elimination is equality interpolating
\cite{BruttomessoGR14}.

\subsection{Local Theory Extensions} 
\label{local}

Let $\Pi_0 {=} (\Sigma_0, {\sf Pred})$ be a signature, and ${\cal T}_0$ be a 
``base'' theory with signature $\Pi_0$. 
We consider 
extensions $\T := {\cal T}_0 \cup \K$
of ${\cal T}_0$ with new function symbols $\Sigma_1$
({\em extension functions}) whose properties are axiomatized using 
a set $\K$ of (universally closed) clauses 
in the extended signature $\Pi = (\Sigma_0 \cup \Sigma_1, {\sf Pred})$, 
which contain function symbols in $\Sigma_1$. 
If $G$ is a finite set of ground $\Pi^C$-clauses, where $C$ is an
additional set of constants, and $\K$ a set of $\Pi$-clauses, we 
will denote by  ${\sf st}({\cal K}, G)$ (resp.\ ${\sf est}({\cal K}, G)$) the set of all 
ground terms (resp.\ extension ground terms, i.e.\ 
terms starting with a function in $\Sigma_1$) 
which occur in $G$ or ${\cal K}$. In this paper we regard every finite set $G$
of ground clauses as the ground formula $\bigwedge_{C \in G} C$. 
If $T$ is a set of ground terms in the signature  $\Pi^C$, 
we denote by $\K[T]$ the set of all instances of $\K$ in which the terms 
starting with a function symbol in $\Sigma_1$ are in $T$. 
Let $\Psi$ be a map associating with 
every finite set $T$ of ground terms a finite set $\Psi(T)$ of ground
terms containing $T$. 
For any set $G$ of ground $\Pi^C$-clauses we write 
$\K[\Psi_{\cal K}(G)]$ for $\K[\Psi({\sf est}({\cal K}, G))]$.
We define:
 
\medskip
\noindent \begin{tabular}{ll}
${\sf (Loc}_f^\Psi)$~ &  For every finite set $G$ of ground clauses in
$\Pi^C$ it holds that\\
&  $\T_0 \cup {\cal K} \cup G \models \bot$ if and only if $\T_0
\cup \K[\Psi_{\cal K}(G)] \cup G$ is unsatisfiable. 
\end{tabular}

\medskip
\noindent Extensions satisfying condition ${\sf (Loc}_f^\Psi)$ are called
{\em $\Psi$-local}. 
If $\Psi$ is the identity 
we obtain the notion of {\em local theory  
extensions} \cite{sofronie-cade-05}; if in addition $\T_0$ is
the theory of pure equality we obtain the notion of 
{\em local theories}  \cite{McAllester93,Ganzinger-01-lics}.

\smallskip
\noindent {\bf Hierarchical reasoning.}
Consider a $\Psi$-local theory extension 
${\cal T}_0 \subseteq {\cal T}_0 \cup {\cal K}$.
Condition $({\sf Loc}_f^{\Psi})$ requires that for every finite set $G$ of ground 
$\Pi^C$-clauses, ${\cal T}_0 \cup {\cal K} \cup G \models \perp$ iff 
${\cal T}_0 \cup {\cal K}[\Psi_{\cal K}(G)] \cup G \models \perp$.
In all clauses in ${\cal K}[\Psi_{\cal K}(G)] \cup G$ the function 
symbols in $\Sigma_1$ only have ground terms as arguments, so  
${\cal K}[\Psi_{\cal K}(G)] {\cup} G$ can be flattened 
and purified
by introducing, in a bottom-up manner, new  
constants $c_t \in C$ for subterms $t {=} f(c_1, \dots, c_n)$ where $f {\in}
\Sigma_1$ and $c_i$ are constants, together with 
definitions $c_t {=} f(c_1, \dots, c_n)$. 
We thus obtain a set of clauses ${\cal K}_0 {\cup} G_0 {\cup} {\sf Def}$, 
where ${\cal K}_0$ and $G_0$ do
not contain $\Sigma_1$-function symbols and ${\sf Def}$ contains clauses of the form 
$c {=} f(c_1, \dots, c_n)$, where $f {\in} \Sigma_1$, $c, c_1, \dots,
c_n$ are constants.
\begin{thm}[\cite{sofronie-cade-05,ihlemann-jacobs-sofronie-tacas08,sofronie-ihlemann-10}]
Let ${\cal K}$ be a set of clauses. 
Assume that 
${\cal T}_0 \subseteq {\cal T}_0 \cup {\cal K}$ is a 
$\Psi$-local theory extension. 
For any finite set $G$ of flat ground clauses (with no nestings of
extension functions), 
let ${\cal K}_0 \cup G_0 \cup {\sf Def}$ 
be obtained from ${\cal K}[\Psi_{\cal K}(G)] \cup G$ by flattening and purification, 
as explained above. 
Then the following are equivalent to ${\cal T}_0 \cup {\cal K} \cup G \models \perp$: 
\begin{itemize}
\item[(i)] ${\cal T}_0 {\cup} {\cal K}[\Psi_{\cal K}(G)] {\cup} G \models
  \perp.$ 
\item[(ii)] ${\cal T}_0 \cup {\cal K}_0 \cup G_0 \cup {\sf Con}_0 \models \perp,$ where 

${\sf Con}_0  = \displaystyle{\{ {\bigwedge}_{i =
      1}^n \!\! c_i
    {\approx} d_i {\rightarrow} c {\approx} d \, {\mid}
f(c_1, \dots, c_n) {\approx} c, f(d_1, \dots, d_n) {\approx} d  {\in}
{\sf Def}  \}}.$ 
\end{itemize}
\label{lemma-rel-transl}
\end{thm} 
\noindent  {\bf Locality and embeddability.} 
When establishing links between locality and embeddability we require 
that the clauses in $\K$
are \emph{flat} 
and \emph{linear} w.r.t.\ $\Sigma_1$-functions.
When defining these notions we distinguish between ground and 
non-ground clauses: 
An {\em extension clause $D$ is flat} 
when all symbols 
below a $\Sigma_1$-function symbol in $D$ are variables. 
$D$ is \emph{linear}  if whenever a variable occurs in two terms of $D$
starting with $\Sigma_1$-functions, the terms are equal, and 
no term contains two occurrences of a variable.
A {\em ground clause $D$ is flat} if all symbols below a $\Sigma_1$-function 
in $D$ are constants.
A {\em ground clause $D$ is linear} if whenever a constant occurs in
two terms in $D$ whose root symbol is in $\Sigma_1$, the two terms are identical, and
if no term which starts with a $\Sigma_1$-function contains two occurrences of the same constant.

\smallskip
\noindent 
In \cite{sofronie-ihlemann-10} we showed that for extensions with sets
of flat and linear clauses $\Psi$-locality can be
checked by checking whether a weak embeddability condition of partial into
total models holds (for details see Appendix~\ref{app:local}). The proof
in \cite{sofronie-ihlemann-10} can be extended to situations in which 
the clauses in $\K$ are not linear if a suitable closure operator is 
used, cf.\ also  \cite{sofronie-fuin-2017,peuter-sofronie-cade-2023}. 

\smallskip
\noindent {\bf Examples of local theory extensions.} 
Using the characterization of locality in terms of weak embeddability of 
partial models into total models, 
in \cite{sofronie-cade-05,Sofronie-Ihlemann07,sofronie-ihlemann-ismvl-07,sofronie-lmcs08} 
we gave several examples of local theory extensions: 
\begin{enumerate}
\item Any extension $\T_0 \cup {\sf UIF}_{\Sigma}$ of a theory $\T_0$
  with uninterpreted functions in a set $\Sigma$; 
\item Extensions of partially ordered 
theories 
with a function $f$ satisfying: 

\smallskip
$ ({\sf Mon}_f) \quad \quad  x_1 \leq y_1 \wedge \dots
\wedge  x_n \leq y_n  \rightarrow f(x_1, \dots, x_n) \leq f(y_1,
\dots, y_n).$

\smallskip
\item Extensions of any theory ${\mathcal T}_0$ containing a predicate
  symbol $\leq$ which is reflexive with functions satisfying boundedness 
$({\sf Bound}^t(f))$ or guarded boundedness $({\sf GBound}^t(f))$ conditions

\smallskip
$({\sf Bound}^t(f)) \quad  \quad \forall x_1, \dots, x_n (f(x_1, \dots, x_n) \leq t(x_1, \dots, x_n))$

$({\sf GBound}^t(f)) \quad \forall x_1, \dots, x_n (\phi(x_1, \dots, x_n) \rightarrow f(x_1, \dots, x_n) \leq t(x_1, \dots, x_n)),$

\smallskip
\noindent 
where $t(x_1, \dots, x_n)$ is a term in the base signature $\Pi_0$ and 
$\phi(x_1, \dots, x_n)$ a conjunction of literals in the signature $\Pi_0$, 
whose variables are in $\{ x_1, \dots, x_n \}$.

\smallskip
\item Extensions of 
 the theory ${\sf
    TOrd}$ of totally ordered sets, $LI({\mathbb R})$ or the theory of
  real numbers 
with functions satisfying 
${\sf SGc}(f,g_1, \dots, g_n) \wedge {\sf Mon}(f, g_1, \dots, g_n)$.

\smallskip
$({\sf SGc}(f,g_1, \dots, g_n)) \quad \forall x_1,\dots, x_n, x ( \bigwedge_{i = 1}^n x_i  \leq g_i(x) \rightarrow f(x_1, \dots, x_n) \leq x)$
\end{enumerate}

\noindent {\bf Chains of theory extensions.} 
In many cases ${\cal K} = {\cal K}_1 \cup
\dots \cup {\cal K}_n$ and we have a chain of local theory
extensions 
$ {\cal T}_0 \subseteq {\cal T}_0 \cup {\cal K}_1 \subseteq  \dots
\subseteq {\cal T}_0 \cup {\cal K}_1 \cup \dots \cup {\cal K}_n,$
with the property that for every extension all variables occur also
below extension functions. In this case checking the satisfiability of 
a set of ground clauses $G$ can be reduced in $n$ steps to a 
satisfiability test w.r.t.\ ${\cal T}_0$. 
Also in this case the number of instances can be expressed using 
a closure operator $\Psi$.

\smallskip
\noindent {\bf Locality transfer results.} In
\cite{sofronie-ihlemann-10} we analyzed 
the way locality results can be transferred to combinations of
theories. 

\begin{thm}[\cite{sofronie-ihlemann-10}]
\label{thm:combine}
Let $\T_0$ be a $\forall\exists$ theory in the signature $\Pi_0$, and
let $\Sigma_1$ and
$\Sigma_2$ be two disjoint sets of fresh function symbols. 
Let $\K_i$ be a set of universally closed flat and linear clauses 
$\Pi_0 \cup \Sigma_i$-clauses for $i = 1,2$, in which all variables
occur below extension functions.  If the extensions $\T_0
\subseteq \T_0 \cup \K_i$ are local for $i = 1,2$ then the extension 
$\T_0 \subseteq \T_0 \cup \K_1 \cup \K_2$ is local. 
\end{thm}

 \section{Uniform interpolation} 
\label{unif-int}

In \cite{ghilardi-superposition} 
a notion of quantifier-free uniform interpolation w.r.t.\ a
theory related also to a notion of covers introduced in
\cite{Gulwani-Musuvathi}  was defined. We summarize the results 
on uniform quantifier-free interpolation from 
\cite{ghilardi-superposition,ghilardi-dag} and
\cite{ghilardi-combinations} which are relevant for this paper.
\begin{defi}[Uniform interpolation \cite{ghilardi-superposition}]
Let ${\cal T}$ be a theory and $\exists {\overline e} \phi({\overline
  e}, {\overline y})$ an existential formula with free variables
${\overline y}$. 
A quantifier-free formula $\psi({\overline y})$ is a {\em ${\cal
    T}$-cover} (or a ${\cal T}$-uniform interpolant)  
of $\exists {\overline e} \phi({\overline  e}, {\overline y})$ iff 
\begin{itemize}
\item[(1)] 
${\cal T} \models \phi({\overline  e}, {\overline y}) \rightarrow \psi({\overline  y})$, and 

\item[(2)] for every quantifier-free formula $\theta({\overline
  y}, {\overline z})$ with 
${\cal T} \models \phi({\overline
  e}, {\overline y}) \rightarrow \theta({\overline
  y}, {\overline z})$ we have: 
${\cal T} \models \psi({\overline y}) \rightarrow \theta({\overline
  y}, {\overline z})$. 
\end{itemize}
A theory ${\cal T}$ is said to have uniform quantifier-free interpolation iff
every existential formula $\exists {\overline e}
\phi({\overline  e}, {\overline y})$ has a ${\cal T}$-cover. 
\label{cover}
\end{defi}
\begin{thm}[\cite{ghilardi-superposition}] 
A quantifier-free formula $\psi({\overline y})$ is a ${\cal T}$-cover 
of $\exists {\overline e} \phi({\overline  e}, {\overline y})$ iff
condition (1) in Definition~\ref{cover} and condition (2') below hold: 
\begin{itemize}
\item[(2')] for every model $\A$ of ${\cal T}$ and every tuple of
elements in the universe of $\A$, ${\overline a}$, such that 
$\A \models \psi({\overline a})$ there exists another model
$\B$ of ${\cal T}$ such that $\A$ embeds into $\B$ and 
$\B \models \exists {\overline e}
\phi({\overline  e}, {\overline a})$.
\end{itemize}
\label{ghilardi-emb}
\end{thm}

\begin{rem}
{\em We can equivalently formulate the definition of a $\T$-uniform interpolant by
considering $\Pi^C$-formulae $\phi( {\overline e}, c_1,\dots,c_n)$. 
A $\T$-uniform interpolant of $\phi$ w.r.t.\ $C_s = \{ c_1, \dots, c_n \}$ is a $\Pi^C$-formula
$\psi(c_1,\dots,c_n)$ which contains only constants in $\{ c_1, \dots,
c_n\}$, and such that 
\begin{itemize}
\item[(1)] $\phi \models_{\T} \psi$, and 
\item[(2)] for every 
formula $\theta$ having only constants in $\{ c_1, \dots, c_n \}$
in common with $\phi$, if $\phi \models_{\T} \theta$ then $\psi
\models_{\T} \theta$. 
\end{itemize}
With this change in notation, condition (2') in
Theorem~\ref{ghilardi-emb} can be formulated as follows: 
\begin{itemize}
\item[(2')] for every $\Pi^C$-structure $\A$ which is a model of ${\cal T}$ 
such that $\A \models \psi(c_1,\dots,c_n)$ there exists another model
$\B$ of ${\cal T}$ such that $\A_{|\Pi}$ embeds into $\B_{|\Pi}$ and the constants
$c_1, \dots, c_n$ have the same interpretation in $\A$ and in $\B$
such that $\B \models \phi( {\overline e}, c_1, \dots, c_n)$.
\end{itemize}}
\end{rem}

\medskip 
\noindent In \cite{ghilardi-superposition} it was proved that the theory ${\sf
  UIF}_{\Sigma}$ of
uninterpreted function symbols in a set $\Sigma$ allows uniform
quantifier-free interpolation; a hierarchical superposition
calculus is used for computing uniform interpolants (covers). 
DAG-methods and tableau methods which allow the computation of 
uniform interpolants are proposed in \cite{ghilardi-dag}. 
The methods in \cite{ghilardi-superposition,ghilardi-dag} always
compute an ${\sf UIF}_{\Sigma}$-uniform interpolant for a formula
$\phi$ containing only function symbols in $\Sigma$ which occur in
$\phi$. 
\begin{lem}
Let $\Pi = (\Sigma, \emptyset)$, $\phi$ be a $\Pi^C$-formula and 
$C_s \subseteq C$ be a set of constants occurring in $\phi$. 
Then there is a ${\sf UIF}_{\Sigma}$-uniform interpolant $\psi$ of $\phi$ w.r.t.\
$C_s$ which contains only function symbols occurring in $\phi$. 
\label{unif-int-no-new-functions}
\end{lem}

\smallskip
\noindent In \cite{ghilardi-combinations} it was shown that a universal theory
${\cal T}$ has a model completion ${\cal T}^*$ iff ${\cal T}$ has
uniform quantifier-free interpolation and, moreover, 
situations were identified 
in which the combination of two theories with uniform 
quantifier-free interpolation has again uniform quantifier-free interpolation.
\begin{thm}[\cite{ghilardi-combinations}]
Let ${\cal T}_1, {\cal T}_2$ be two convex, stably infinite, equality
interpolating, universal theories over disjoint signatures admitting 
uniform quantifier-free interpolation. Then ${\cal T}_1 \cup {\cal  T}_2$ admits uniform quantifier-free interpolation. 
\label{thm-comb}
\end{thm}

\begin{ex}
The theory of uninterpreted function symbols ${\sf
  UIF}_{\Sigma}$ is universal, convex, stably infinite and
  equality interpolating \cite{BruttomessoGR14} and admits uniform quantifier-free
  interpolation \cite{ghilardi-superposition}.
Therefore for every theory $\T$ with signature disjoint from $\Sigma$
which is convex, stably infinite, equality
interpolating, and universal, if $\T$ admits uniform quantifier-free
interpolation then $\T \cup {\sf UIF}_{\Sigma}$ admits uniform quantifier-free
interpolation.
 
Moreover, for every formula $\phi$ we can compute a $\T \cup {\sf
  UIF}_{\Sigma}$-uniform interpolant containing only symbols in the
signature of $\T$ and extension symbols occurring in $\phi$. 

This holds in particular when $\T = LI({\mathbb R})$  (linear real
arithmetic).  $LI({\mathbb R})$ has a universal
  axiomatization, is convex, stably infinite, allows quantifier
  elimination, hence (i) has also the equality interpolation property
  \cite{BruttomessoGR14} and (ii) allows uniform quantifier-free interpolation.
Therefore $LI({\mathbb R}) \cup {\sf  UIF}_{\Sigma}$ admits uniform
quantifier-free interpolation.
For computing the uniform interpolant of a formula $\phi$, 
the convex combined cover algorithm proposed in
\cite{ghilardi-combinations} can be used
for computing a $LI({\mathbb R}) \cup {\sf
  UIF}_{\Sigma}$-interpolant of $\phi$. This uniform interpolant
contains only extension symbols in $\Sigma$ occurring in $\phi$. 
\label{ex:lir-uif}
\end{ex}

\section{General uniform interpolation} 
\label{gen-unif-int}
\label{sect:gen-unif-int} 
We define the notion of general uniform quantifier-free
interpolation, and show that the semantic characterization in Theorem~\ref{ghilardi-emb}
(proved in \cite{ghilardi-superposition})  
can be extended to the case of  general uniform quantifier-free
interpolation. 

\smallskip
\noindent Let $\Pi = (\Sigma, {\sf
  Pred})$ be a signature where $\Sigma = \Sigma_0 \cup \Sigma_1$, where 
$\Pi_0 = (\Sigma_0, {\sf  Pred})$ is a set of interpreted function and
predicate symbols and $\Sigma_1$ a set of uninterpreted
function symbols. Let $\phi$ be a $\Pi^C$-formula
and let $\Sigma_s \subseteq \Sigma_1$, $C_s \subseteq C$  be 
sets of uninterpreted symbols and constants occurring in $\phi$. 
We denote by $\Pi_s$ the signature $(\Sigma_0 \cup \Sigma_s, {\sf Pred})$. 
We denote by $\Sigma_e$ the set of uninterpreted 
function symbols in $\phi$ which are not in $\Sigma_s$ and 
by $C_e$ the set of additional constants in $C$ occurring in $\phi$ which
are not in  $C_s$ (the symbols which are eliminated).
Let $\Pi_r = (\Sigma_0 \cup (\Sigma_1 \backslash
  \Sigma_e), {\sf Pred})$ and  
$C_r = C \backslash C_e$ (the set of function and predicate symbols
and constants in $\Pi^C$ which are not eliminated). 
\begin{defi}[General uniform interpolation]
Let ${\mathcal T}$ be a theory over a signature $\Pi = (\Sigma, {\sf
  Pred})$, where $\Sigma = \Sigma_0 \cup \Sigma_1$, where 
$\Pi_0 = (\Sigma_0, {\sf  Pred})$ is a set of interpreted function
symbols and $\Sigma_1$ a set of uninterpreted
function symbols. 
We say that ${\mathcal T}$ has the {\em general uniform interpolation 
property} if for every quantifier-free $\Pi^C$-formula $\phi$
(where $C$ is a set of new constants) and all subsets
$\Sigma_s \subseteq \Sigma_1$, $C_s \subseteq C$  
of uninterpreted symbols and constants occurring in $\phi$ 
there exists a quantifier-free $\Pi_s^{C_s}$-formula 
$\psi$ such that: 
\begin{itemize}
\item[(1)] $\phi \models_{\cal T} \psi$, and 
\item[(2)] For every ground formula $\theta$ having in common
  with $\phi$ only interpreted symbols in $\Pi_0$, uninterpreted
  symbols in $\Pi_s$, and constants in $C_s$ if 
$\phi \models_{\cal T} \theta$ then $\psi \models_{\cal T} \theta$.
\end{itemize} 
A formula $\psi$ with the properties (1) and (2) is called a ${\cal
  T}$-general uniform interpolant of $\phi$ w.r.t.\ $\Sigma_s
\cup C_s$. 
\label{def-general-unif-int}
\end{defi}
\begin{rem}
Definition~\ref{cover} can be seen as a special case of
Definition~\ref{def-general-unif-int}, in which  
all function and predicate symbols in the signature of
$\T$ are {\em interpreted}, i.e.\ can be contained in the 
uniform interpolant of a formula.

We will later also consider the possibility of regarding symbols in 
$\Sigma_1$ as ``semi-interpreted'' in the following sense: 
If two function symbols $f, g \in \Sigma_1$ occur together in an 
axiom of ${\cal T}$, and $f \in \Sigma_s$ then the general uniform
interpolant is allowed to also contain the function symbol $g$.
We will model this using a closure operator $\Theta$, depending on the theory
${\cal T}$ and require that the general uniform interpolant of a
formula $\phi$ can contain any
$\Pi_0$-symbols and only uninterpreted symbols in $\Theta(\Sigma_s \cup C_s)$, where
$\Sigma_s \cup C_s$ are 
uninterpreted function symbols and constants occurring in the formula $\phi$.
\label{remark} 
\end{rem}
\begin{thm}
Let $\T$ be a theory with signature $\Pi = (\Sigma_0 \cup \Sigma, {\sf  Pred})$, 
and let $\phi$ be a ground
$\Pi^C$-formula. Let $\Sigma_s \subseteq \Sigma$ be a subset of the
uninterpreted function symbols occurring in $\phi$, and $C_s \subseteq C$ a subset of the
constants occurring in $\phi$. We use the notation introduced in the
paragraph preceding 
Definition~\ref{def-general-unif-int}.
 
\noindent A ground $\Pi_s^{C_s}$-formula $\psi$ 
is a ${\cal T}$-general uniform interpolant of $\phi$  w.r.t.\ $\Sigma_s \cup
C_s$ if and only if the following conditions hold:
\begin{itemize}
\item[(1)]  ${\cal T} \models \phi  \rightarrow \psi$, and 
\item[(2')] for every $\Pi^C$-structure 
$\A$ such that $\A$ is a model of ${\cal T}$
and $\A \models \psi$ there exists another model
$\B$ of ${\cal T}$ such that 
$\A_{|\Pi_r^{C_r}}$ embeds into $\B_{|\Pi_r^{C_r}}$ and 
$\B \models \phi$.
\end{itemize}
\vspace{-2mm}
\label{gen-un-int-sem}
\end{thm}
{\em Proof:} 
Assume that $\psi$ satisfies conditions (1)
and (2') above. 
We show that for every  quantifier-free formula $\theta$ having in common
  with $\phi$ only uninterpreted symbols and constants in
  $\Sigma_s \cup C_s$, if 
$\phi \models_{\cal T} \theta$ then $\psi \models_{\cal T} \theta$.
Let $\A$ be a model of ${\cal T}$ and of $\psi$. 
Then, by condition (2'), there exists another model
$\B$ of ${\cal T}$ such that 
$\A_{|\Pi_r^{C_r}}$ embeds into $\B_{|\Pi_r^{C_r}}$ and 
$\B \models \phi$.
As $\phi \models_{\cal T} \theta$, $\B \models \theta$. 
Since $\theta$ does not contain any symbol in $\Sigma_e \cup C_e$, 
it is a $\Pi_r^{C_r}$-formula, so $\B_{|\Pi_r^{C_r}} \models \theta$. 
Since $\A_{|\Pi_r^{C_r}}$ is a substructure of ${\cal  B}_{|\Pi_r^{C_r}}$
and $\theta$ is a ground formula, $\A_{|\Pi_r^{C_r}}\models
\theta$, 
so $\A \models \theta$. 

Assume now that $\psi$ is a uniform interpolant of
$\phi$ w.r.t.\ $\Sigma_s \cup C_s$. 
Then (1) holds. We show that (2') holds as well. 
Let $\A$ be a model of ${\cal T}$ 
in which $\psi$ holds.  Since $\psi$ does not contain symbols in
$\Sigma_e \cup C_e$, it follows that  $\A_{|\Pi_r^{C_r}}\models \psi$. 
To construct a  model $\B$ of ${\cal T}$ with the desired
properties we proceed as follows: 
We prove that $\Delta(\A_{|\Pi_r^{C_r}}) \cup \phi$ is a 
${\cal  T}$-consistent set of sentences. 
Assume that this is not the case. By compactness, there is a finite 
subset $A = \{ l_1, \dots, l_n \}$ of literals in $\Delta(\A_{|\Pi_r^{C_r}})$ 
such that $A \cup \phi \models_{\cal T} \perp$, i.e.\ 
$\phi \models_{\cal T} \neg l_1 \vee \dots \vee \neg l_n$.
The uninterpreted function symbols and constants shared by 
$\theta := \neg l_1 \vee \dots \vee \neg l_n$ and $\phi$ are contained 
in $\Sigma_s \cup C_s$. Therefore, by the definition of a uniform
interpolant, $\psi \models_{\cal T} \theta$.
We know that $\A \models \psi$. 
It would therefore follow that $\A \models \neg l_1 \vee \dots
\vee \neg l_n$, which is a contradiction. 
It follows that $\Delta(\A_{|\Pi_r^{C_r}}) \cup \phi$ is a ${\cal
  T}$-consistent set of sentences, so there exists a model of 
${\cal T}$, ${\cal B}$, such that  $\A_{|\Pi_r^{C_r}}$ embeds into
 ${\cal B}_{|\Pi_r^{C_r}}$ 
and $\B \models \phi$. \QED

\section{General uniform interpolation in theory extensions}
\label{symbol-elimination}
We present a method for eliminating function symbols based on an
adaptation of a method for symbol elimination in theory extensions 
we proposed in \cite{sofronie-lmcs18}. 
In \cite{sofronie-lmcs18} we considered theory
extensions $\T_0 {\cup} \K$ with extension symbols $\Sigma$, 
and assumed that the function symbols in a subset 
$\Sigma_s \subseteq \Sigma$ were  underspecified, i.e.\ considered to be parameters. 
Given a set $G$ of  ground clauses such that $\T_0 {\cup} \K {\cup} G$ is
satisfiable, we used a form of symbol elimination 
for deriving a (weakest) universally quantified condition $\Gamma$ on the
parameters in $\Sigma_s$ such that $G$ becomes unsatisfiable w.r.t. 
$\T {\cup} (\K {\cup} \Gamma)$. 
This was used in verification, for deriving additional conditions 
under which certain properties are guaranteed to be inductive
invariants \cite{sofronie-fundamenta20} and
for invariant strengthening \cite{peuter-sofronie19}. 

\medskip
\noindent We show that the method can be used for 
reducing the problem of computing a general uniform interpolant in 
local theory extensions to the problem of computing uniform interpolants.

\medskip
\noindent Let $\Pi_0 = (\Sigma_0, {\sf Pred})$. 
Let ${\mathcal T}_0$ be a base theory with signature $\Pi_0$.  
We consider theory extensions $\T_0
\subseteq \T = \T_0 \cup \K$ with signature $\Pi = \Pi_0 \cup \Sigma$ 
and sets of ground clauses $G$ over $\Pi^C$. 
Among the extension functions $\Sigma$ (resp.\ constants $C$)  
we identify sets $\Sigma_s$  (resp.\  $C_s$)  of function (resp.\
constant)  symbols which
are considered to be shared and must be kept, and the set $\Sigma_1 =
\Sigma \backslash \Sigma_s$. \\
We will proceed as follows: 
\begin{description}
\item[(I)]   We first eliminate the function symbols in $\Sigma_1$
  with arity $\geq$ 1 and the constants which do not occur below 
functions in $\Sigma_s$. 
\item[(II)]  We then use (if possible) methods similar to those
  developed in \cite{ghilardi-combinations} to eliminate the remaining
  constants which are not in $\Sigma_s \cup C_s$.
\end{description}

\begin{algorithm}[t]
\caption{Algorithm for Function Elimination}

\begin{tabular}{ll}
{\bf Input:} & \!\!\!\!Theory extension ${\cal T}_0 \subseteq {\cal T}_0 \cup
{\cal K}$ with signature $\Pi = \Pi_0 \cup (\Sigma_s \cup \Sigma_1)$\\
& ~~~where ${\cal K}$ is a finite set of flat and linear $\Pi$-clauses; \\
& \!\!\!\!$G$, a finite set of flat and linear ground clauses in the signature $\Pi^C$; \\
& \!\!\!\!$T$, a finite set of flat ground $\Pi^C$-terms  s.t.\ ${\sf
  est}({\cal K}, G) \subseteq T$ and ${\cal K}[T]$ is ground. \\

{\bf Output:} & Ground  $\Pi_s^{C}$-formula $\Gamma$. \\
\hline 
\end{tabular}

\begin{description}
\vspace{-1mm}
\item[Step 1] Compute the set of $\Pi_0^C$-clauses $\K_0 {\cup} G_0 {\cup}
  {\sf Con}_0$  from $\K[T] \cup G$  using purification
  (cf.\ Thm.~\ref{lemma-rel-transl}, extension
  symbols $\Sigma = \Sigma_s \cup \Sigma_1$).

\item[Step 2]  $G_1 := {\mathcal K}_0 \cup G_0\cup {\sf Con}_0$. 
Among the constants in $G_1$, identify 
\begin{enumerate}
\item[(i)] the constants
${\overline c_f}$, $f {\in} \Sigma_s$, where $c_f {=} f {\in}
\Sigma_s$ is a constant
parameter or $c_f$ is 
introduced by a definition $c_f {:=} f(c_1,\dots,c_k)$ in the hierarchical
reasoning method, 
\item[(ii)] all constants  ${\overline c_p}$ 
occurring as arguments of functions in $\Sigma_s$ in such definitions. 
\end{enumerate}
Let ${\overline  c}$ be the remaining constants.\\
Replace the constants in ${\overline  c}$
with existentially quantified variables ${\overline x}$ in $G_1$,
i.e.\ 
replace $G_1({\overline c_p}, {\overline c_f}, {\overline c})$ 
with $G_1({\overline c_p}, {\overline c_f}, {\overline x})$, and
consider the formula
$\exists {\overline x} G_1({\overline c_p},{\overline c_f}, {\overline x})$.


\item[Step 3] Compute a quantifier-free formula 
$\Gamma_1({\overline c_p}, {\overline c_f})$  equivalent to 
$\exists {\overline x} G_1({\overline c_p}, {\overline c_f},{\overline
  x})$ w.r.t.\ $\T_0$ using a  method for quantifier elimination in 
${\mathcal T}_0$.  


\item[Step 4] Let $\Gamma({\overline c_p})$ be the formula 
obtained by replacing back in $\Gamma_1({\overline c_p}, {\overline c_f})$ 
the constants $c_f$ introduced by definitions $c_f := f(c_1, \dots,
c_k)$ with the terms $f(c_1, \dots,c_k)$. 
\vspace{-1mm}
\end{description}
\label{alg-symb-elim}
\end{algorithm}

\subsection{(I) Eliminating function symbols in local theory
  extensions} 
\label{elim-f-symbols}
We identify situations in which we can
generate, for every set of flat ground clauses $G$,   
a ground formula $\Gamma$ containing only uninterpreted symbols in 
$\Sigma_s$, such that $\Gamma$ is a general uniform interpolant of $G$
w.r.t.\ $\Sigma_s \cup {\overline
  C_s}$, 
for a subset ${\overline C}_s$ of $C$ containing the constants $C_s$ in $G$.
We will use a method which eliminates symbols in a hierarchical way, by reducing 
the problem to quantifier elimination in the theory ${\cal T}_0$
which is described in Algorithm~\ref{alg-symb-elim}. 

\medskip
\noindent We first consider a special type of extensions, in
which $\K$ is the disjoint union between a set of clauses $\K_s$ 
in which only symbols in $\Pi_0 \cup \Sigma_s$ 
occur, a set of clauses $\K_1$ in which only the symbols in $\Pi_0 \cup \Sigma_1$ 
(to be eliminated) occur, and a set of clauses $\K_i$ containing only
symbols in $\Pi_0$ and 
extension symbols not occurring in $G$ (and therefore irrelevant for $G$) are disjoint.

\begin{thm}
Assume that ${\mathcal T}_0$ allows quantifier elimination. 
Let ${\mathcal T}_0 \subseteq {\mathcal T}_0 \cup \K$ be an extension
of the theory ${\mathcal T}_0$ with additional function symbols in a
set $\Sigma$ which is the disjoint union of $\Sigma_s$, $\Sigma_1$ and 
$\Sigma_i$,  satisfying a finite set $\K$ of flat
and linear
clauses such that for every clause in $\K$, every variable occurs below
an extension symbol.
Let $G$ be a finite set of flat and linear
ground $(\Pi_0 \cup \Sigma_s \cup \Sigma_1)^C$-clauses 
and $T$ a set of
flat $(\Pi_0 \cup \Sigma_s \cup \Sigma_1)^C$-terms satisfying the conditions in Algorithm~1, and let
$\Gamma_T$ 
be the ground formula obtained by applying Algorithm~1.
\begin{itemize}
\item[(1)] $G \models_{{\cal T}_0 \cup {\cal K}} \Gamma_T$.
\item[(2)] 
Assume that ${\cal T}_0 \subseteq {\cal T}_0 \cup {\cal K}$ is a local
theory extension, and that ${\cal K} = {\cal K}_s \cup {\cal K}_1 \cup
{\cal K}_i$,
where ${\cal K}_s$ is a set of $\Pi_0 \cup \Sigma_s$-clauses, ${\cal K}_1$ is a
set of $\Pi_0 \cup \Sigma_1$-clauses, and ${\cal K}_i$ is a set of $\Pi_0 \cup \Sigma_i$-clauses.
Let ${\overline C}_s$ consist of $C_s$ together with all constants in
$C$ which occur as arguments of functions in
$\Sigma_s$ in $G$. 
Then the formula $\Gamma$ obtained by 
applying Algorithm 1\footnote{Since $G$ does not contain the symbols
  in $\Sigma_i$, the instances computed in Step 1 of Algorithm 1 do
  not contain instances of $\K_i$, so $\K_i$ can be ignored.} 
to $\T_0 \cup \K_s \cup \K_1$, $G$ and $T := {\sf  est}(\K, G)$ 
is a $\T_0 \cup \K$-general uniform interpolant of $G$ w.r.t.\
$\Sigma_s \cup {\overline C}_s$.
\end{itemize}
\vspace{-3mm}
\label{symb-elim-simplif}
\label{inv-trans-qe}
\end{thm}
{\em Proof:} 
Assume that $\Sigma_1$ and $C_e$ are the function symbols  
and constants in $C$ occurring in $G$ which are not in 
$\Sigma_s$ or ${\overline C}_s$. 
We use the notation: $\Pi_s = (\Sigma_0 \cup \Sigma_s, {\sf Pred})$, 
$\Pi_s = (\Sigma_0 \cup \Sigma_s, {\sf Pred})$, and
$C_r = C \backslash C_e$, $\Pi_r = \Pi_s \cup \Sigma_i$. 

Note first that because $G$ is a set of flat ground clauses and ${\cal K}$ is
a set of flat clauses, the only arguments of function symbols in
$\Sigma$ in ${\cal K}[G] \cup G$ are constants in $\Sigma \cup C$. 
Moreover, $G$ and ${\cal K}[G]$, hence also ${\sf Def}$,
only contain extension terms occurring in $G$ (thus they do not
contain function symbols in $\Sigma_i$). 
Since $G$ does not contain function symbols in $\Sigma_i$, $\K_r[G] =
\emptyset$, so $\K[G] = \K_s[G] \cup \K_1[G]$. 

By the way in which  the set of constants 
${\overline c}$ to be eliminated in Step 3 of Algorithm 1 is chosen,
it does not contain any 
constant in ${\overline C}_s$. Therefore, the formula $\Gamma$
obtained in Step 4 of Algorithm~1 is a $\Pi_s^{{\overline
    C}_s}$-formula. 
We show that it has the properties of a $\T_0 \cup \K$-general uniform interpolant of $G$ w.r.t.\
$\Sigma_s \cup {\overline C}_s$. 

\smallskip
\noindent (1) We show that $G \models_{\T_0 \cup \K} \Gamma$. Let $\A$ be a $\Pi^C$-structure which is a 
model of ${\cal T}_0 \cup {\cal K}$ in which 
$G$ holds. Let $\A^{C'}$ be
the expansion of $\A$ with the constants $C'$ introduced in Step 1
of Algorithm 1, interpreted according to the definitions ${\sf Def}$.
Then $\A^{C'}$ is a model for ${\cal K}_0 \cup G_0 \cup {\sf
  Con}_0 \cup {\sf Def}$, so 
$\A^{C'}_{|\Pi_0^{C \cup C'}}$ is a model for ${\cal
  K}_0 \cup G_0 \cup {\sf Con}_0$, thus (with the notations used in Algorithm 1) 
$\A^{C'}_{|\Pi_0^{C \cup C'}}$ is a model for 
$\exists {\overline c} ({\cal K}_0 \cup G_0 \cup {\sf Con}_0)$. 
We know that $\exists {\overline c} ({\cal K}_0 \cup G_0 \cup {\sf Con}_0)$ is equivalent 
w.r.t.\ ${\cal T}_0$ to the formula $\Gamma_1$ obtained in Step 3. 
Therefore, $\A^{C'}_{|\Pi_0^{C \cup C'}} \models \Gamma_1$.
As $\A^{C'}$ is a model for ${\sf Def}$, 
it follows that $\A$ is a model of $\Gamma$. 

\smallskip
\noindent (2) 
Let $\theta$ be a ground formula having in common with $G$ only
symbols in $\Pi_s^{{\overline C}_s}$ such that 
$G \models_{{\cal T}_0 \cup {\cal K}} \theta$. 
We show that $\Gamma \models_{{\cal  T}_0 \cup {\cal K}} \theta$ 
(more precisely, that condition (2') in Theorem~\ref{gen-un-int-sem}  holds). 
Let $\A$ be a $\Pi^C$-structure which is a model of 
${\cal T}_0 \cup {\cal K}$ and such that $\A \models \Gamma$.
If we purify $\Gamma$ by introducing new constants\footnote{W.l.o.g.\
  we can assume that we use the same names used in Algorithm~1.} denoting terms 
starting with function symbols in $\Sigma_s$ we obtain a ground
formula $\Gamma_1 \wedge {\sf Def}_s$. Let $C'$ be the constants introduced in this
process, and $\A^{C'}$ be the extension of ${\cal  A}$ with 
these new constants, defined as described by the definitions ${\sf Def}_s$. 
Then $\A^{C'} \models \Gamma_1 \wedge {\sf Def}_s$, so 
$\A^{C'}_{|\Pi_0^{C'}} \models \Gamma_1$ (which is equivalent 
w.r.t.\ ${\cal T}_0$ to  $\exists {\overline c} ({\cal K}_0 \cup
G_0 \cup {\sf Con}_0)$). 
Therefore, $\A^{C'}_{|\Pi_0^{C'}}  \models \exists {\overline c} ({\cal K}_0 \cup
G_0 \cup {\sf Con}_0)$. 
This shows that there are values ${\overline a}$ for the constants in ${\overline c}$
(perhaps different from the interpretations of these constants in
$\A^{C'}$) for which $({\cal K}_0 \cup
G_0 \cup {\sf Con}_0)$ holds. 
Let $C''$ be $C'$ to which we add the constants ${\overline c}$
determined in Step 2. 
Let ${\overline {\cal A}}$ be a $\Pi^{C \cup C''}$-structure with the
same support as ${\cal A}$, 
which agrees with $\A^{C'}$ on $\Pi_0, \Sigma_s, \Sigma_i$, 
$C_s$ and $C'$,  and such that the interpretation of 
the constants in ${\overline c}$ is ${\overline a}$. 
Then  ${\overline {\cal A}} \models {\cal K}_0 \cup G_0 \cup {\sf Con}_0$. 
Note that the constants in ${\overline c}$ do not occur among the
constants in $C' \cup C_s$ or as arguments of function symbols in
$\Sigma_s$ or $\Sigma_i$, so 
${\overline {\cal A}}$  is still a model of ${\sf Def}_s$, of $\K_s$
and of $\K_i$. 
We construct a partial
$\Pi^C$-algebra ${\cal P}$, with the same support as ${\overline \A}$
where:
\begin{itemize}
\item all symbols in $\Pi_0, \Sigma_s$ and $\Sigma_i$ 
and all constants in  ${\overline c}$ are defined as in ${\overline {\cal A}}$;
\item partial functions in $\Sigma_1$ are defined according to the definitions in
${\sf Def}$. 
\end{itemize}
The definition of partial functions in $\Sigma_s$, as reconstructed from
the definitions, coincides with the interpretation of those function
symbols in $\A$, 
because the constants in ${\overline c}$ do not occur below
function symbols in $\Sigma_s$.
The partial structure ${\cal P}$ satisfies 
${\cal K}[G] \cup G$, where ${\cal K}[G] = {\cal K}_s[G] \cup {\cal K}_1[G]$:
\begin{itemize}
\item The functions
in $\Sigma_0 \cup \Sigma_s \cup \Sigma_i$ 
are totally defined in ${\cal P}$, and they are
defined as in $\A$, which is a model of $\T_0 \cup \K_s \cup \K_i$, so ${\cal P}$ is a
model of $\T_0 \cup \K_s \cup \K_i$.
\item Only the $\Sigma_1$-terms in ${\cal K}_1[G]$ are
defined in ${\cal P}$, and all clauses in $\K_1[G]$ are true in ${\cal
  P}$ because the clauses in $\K_1$ contain only function symbols in
$\Pi_0 \cup \Sigma_1$ and ${\overline A} \models \K_0 \cup {\sf Def}$.
\end{itemize}
Thus,  ${\cal P}$ is a total model of $\T_0 \cup {\cal K}_s \cup \K_i$ 
and it weakly satisfies ${\cal K}_1$, hence it is a weak partial model
of $\T_0 \cup {\cal K}$ with totally defined $\Pi_0$-functions, 
in which all terms in $G$ are defined and $G$ holds. 
By the locality of the extension ${\cal T}_0 \subseteq {\cal T}_0 \cup
{\cal K}$, ${\cal P}$ weakly embeds into a total model $\B$ of ${\cal T}_0
\cup {\cal K}$ which is a model of $G$, hence also
of $\theta$. 
Since in ${\cal P}$ the functions in $\Pi_r = \Pi_0 \cup \Sigma_s \cup
\Sigma_1$ and the constants in
$C_r = C_s$ are totally defined (and are defined as in $\A$) 
and the only functions which are partial are those in $\Sigma_1$, 
${\cal P}_{|\Pi_r^{C_r}} = \A_{|\Pi_r^{C_r}}$, i.e.\ $\A_{|\Pi_r^{C_r}}$ embeds into  $\B_{|\Pi_r^{C_r}}$.
Since $\theta$ is ground, only
contains symbols in the signature $\Pi_r^{C_r}$, 
and holds in $\B$ it also holds in $\A$. \QED

\begin{rem}
{\em 
The proof of Theorem~\ref{inv-trans-qe}(2) uses the fact
that the clauses in $\K_s$ contain only symbols in $\Pi_0^C \cup
\Sigma_s$ and those in $\K_1$  contain only symbols in $\Pi_0^C \cup
\Sigma_1$. The arguments used in the proof can be slighty changed
such that they work also in cases in which the clauses in $\K_1$
are allowed to contain symbols in $\Sigma_s$ under the assumption 
that the extension is $\Psi$-local, for a suitable closure operator
$\Psi$ which guarantees that in the definiton of the partial
structure ${\cal P}$, only the terms in $\K_1[\Psi(G)]$ are defined, 
so ${\cal P}$ is a weak partial model of $\K_1$. In Algorithm~1
$\K[\Psi(G)]$ needs to be used instead of $\K[G]$, and $T = \Psi({\sf
  est}(\K, G))$ (cf. also the proof of
Theorem~\ref{thm:definability} (Claim 2, Step 2)).
}
\end{rem}

\begin{cor}
Let $\T = \T_0 \cup {\sf UIF}_{\Sigma}$ with signature $\Pi$, and let
$G$ be a finite set of flat
and linear ground $\Pi^C$-clauses.  Then the $\Pi_s^{\overline{C_s}}$-formula $\Gamma$ obtained by 
applying Algorithm 1 to $\T$, $G$ and $T := {\sf  est}(\K, G)$ is a
$\T_0 \cup \K$-general uniform interpolant of $G$ w.r.t.\
$\Sigma_s \cup {\overline C}_s$. 
\end{cor}
{\em Proof:} Immediate consequence of Theorem~\ref{inv-trans-qe}, due
to the fact that for extensions with uninterpreted function symbols
all conditions in Theorem~\ref{inv-trans-qe}(2) are fulfilled. \QED

\smallskip
\noindent 
In what follows we will not explicitly refer to
$\Sigma_i$ -- or equivalently assume that it is contained in
$\Sigma_1$.

\begin{prop}
Let $G$ be a finite set
of flat and linear ground $\Pi^C$-clauses, and let $\Sigma_s, \Sigma_1$ be the
extension function symbols occurring in $G$ and $C_s \subseteq C$ a set of constants
occurring in $G$. Let $\T := \T_0 \cup \K_s \cup \K_1 \cup \K_i$ be a
$\Pi$-theory, such that all the conditions in
Theorem~\ref{inv-trans-qe}(2) hold.  
Let $\Gamma$ be the $\T$-general uniform interpolant for $G$ w.r.t.\
$\overline{C_s}$ (with $C_s \subseteq {\overline C_s}$) obtained by applying Algorithm~1. 
If $\psi$ is a $\Pi_s^{C_s}$-ground formula which is a 
$\T_0 \cup \K_s \cup {\sf UIF}_{\Sigma_1 \cup \Sigma_i}$-uniform interpolant of $\Gamma$ w.r.t.\ $C_s$
then $\psi$ is a $\T$-general uniform interpolant of $G$ w.r.t.\ $C_s$.
\label{prop-leave-ax-forgotten-out}
\end{prop}
{\em Proof:} 
We assumed that $\psi$ is a $\Pi_s^{C_s}$-ground formula. 

\smallskip
\noindent (1) By Theorem~\ref{symb-elim-simplif}, $G \models_{\T} \Gamma$. 
Since $\psi$ is a $\T_0 \cup \K_s \cup {\sf UIF}_{\Sigma_1 \cup \Sigma_i}$-uniform interpolant for $\Gamma$ w.r.t.\
  $C_s$, we know that 
$\Gamma \models_{\T_0 \cup \K_s \cup {\sf UIF}_{\Sigma_1 \cup \Sigma_i}} \psi$, so  
  $\Gamma \models_{\T} \psi$. 
Therefore, $G \models_{\T} \psi$.

\smallskip
\noindent (2) We prove that  for every $\Pi^C$-structure 
$\A$ such that $\A$ is a model of ${\cal T}$
and $\A \models \psi$ there exists another model
$\D$ of ${\cal T}$ such that 
$\A_{|\Pi_r^{C_r}}$ embeds into $\D_{|\Pi_r^{C_r}}$ and 
$\D \models G$, where $\Pi_r = \Pi_0 \cup \Sigma_s \cup \Sigma_i$ and
$C_r = C \backslash C_e$,   where
$C_e$ is the set of  all constants in $G$ which are not in $C_s$.  

Let $\A$ be a $\Pi^C$-structure which is a model of ${\cal T}$ and of
$\psi$. 
Then $\A$ is clearly also a model of $\T_0 \cup \K_s \cup \K_i \cup {\sf UIF}_{\Sigma_1}$.

We first show that there exists a $\Pi^C$-structure $\C$ which is a 
model of $\T_0 \cup \K_s \cup \K_i \cup \K_1$  with $\C \models \Gamma$ 
and such that $\A_{|\Pi}$ embeds into $\C_{|\Pi}$
and the interpretation of the constants in $C_s$ is the same in $\A$
as in $\C$.

Since $\psi$ is a $\T_0 \cup \K_s \cup {\sf UIF}_{\Sigma_1 \cup \Sigma_i}$-uniform interpolant of $\Gamma$
w.r.t.\ $C_s$, 
 by Theorem~\ref{ghilardi-emb}, there exists a $\Pi^C$-structure $\B$ which is a
model of $\T_0 \cup \K_s \cup {\sf UIF}_{\Sigma_1 \cup \Sigma_i}$, with $\B
\models \Gamma$ and such that $\A_{|\Pi}$ embeds into $\B_{|\Pi}$
and the interpretation of the constants in $C_s$ is the same in $\A$
as in $\B$. In what follows we regard $\A$ as a $\Pi$-substructure of
$\B$. 

To obtain a model of $\T_0 \cup \K_s \cup \K_i \cup \K_1$ with 
the same properties we proceed as follows:
Let ${\cal P}$ be a partial $\Pi^C$-structure with the same support as
$\B$, in which the $\Pi_0$-function and predicate symbols and all
constants in $C$ are defined 
as in $\B$, and such that (i) all function symbols in $\Sigma_s$ are 
defined as in $\B$, and (ii) for every function symbol $f \in
\Sigma_1 \cup \Sigma_i$: 

\smallskip
$f_{{\cal P}}(b_1, \dots, b_n) = \left\{ \begin{array}{ll} 
f_{\A}(b_1, \dots, b_n) & \text{ if } b_1, \dots, b_n \in |\A| \\
\text{undefined} & \text{ otherwise} 
\end{array} \right.$

\smallskip
\noindent Clearly, ${\cal P}$ is a total model of $\T_0 \cup \K_s$. 
Since $\A$ is a model of $\K_1 \cup \K_i$, and $\K_1$ contains only
symbols in $\Pi_0 \cup \Sigma_1$ and $\K_i$ contains only clauses in
$\Pi_0 \cup \Sigma_i$, ${\cal P}$ is a weak partial model of
$\K_1 \cup \K_i$. 
Therefore, by the locality of the extension  $\T_0  \subseteq \T_0
\cup \K_s \cup \K_i \cup \K_1$,
${\cal P}$ weakly embeds into a total model $\C$ of 
$\T_0 \cup \K_s \cup \K_i \cup \K_1$.

If $f \in \Sigma_s \cup \Sigma_i \cup \Sigma_1$, then by the way
${\cal P}$ is defined and from the fact that ${\cal P}$ weakly embeds
into $\C$ it follows that for all elements $a_1, \dots, a_n \in |\A|$, 
$f_{\A}(a_1, \dots, a_n) = f_{\B}(a_1, \dots, a_n) = f_{{\cal P}}(a_1,
\dots, a_n) = f_{\C}(a_1, \dots, a_n)$. 
In addition, the interpretation of the constants in $C_s$ is the 
same in $\A$, $\B$, ${\cal P}$ and $\C$. 
Since $\B \models \Gamma$ and the definitions of the constants in
$\overline{C_s}$ is the same in $\B$, in ${\cal P}$ and in $\C$, 
$\C \models \Gamma$. 

This shows that  there exists a $\Pi^C$-structure $\C$ which is a
model of $\T_0 \cup \K_s \cup \K_i \cup \K_1$, with $\C
\models \Gamma$ and such that $\A_{|\Pi}$ embeds into
$\C_{|\Pi}$ and  the interpretation of the constants in $C_s$ is the same in $\A$
as in $\C$.

Since $\C$ is a model of $\T := \T_0 \cup \K_s \cup \K_i \cup \K_1$ such that
$\C \models \Gamma$, 
and $\Gamma$ is a $\T$-general uniform interpolant 
of $G$ w.r.t.\ $\Sigma_s \cup \overline{C_s}$, where
$C_s \subseteq \overline{C_s}$, it follows by Theorem~\ref{gen-un-int-sem} 
that there exists a $\Pi^C$ structure $\D$ which is a
model of $\T_0 \cup \K_s \cup \K_i \cup \K_1$, with $\D
\models G$ and such that $\C_{|\Pi_r^{\overline{C_s}}}$ 
embeds into $\D_{|\Pi_r^{\overline{C_s}}}$. 
Since $\A_{|\Pi^{C_s}}$ embeds into
$\C_{|\Pi^{C_s}}$ and $\C_{|\Pi_r^{\overline{C_s}}}$ 
embeds into $\D_{|\Pi_r^{\overline{C_s}}}$ and $C_s \subseteq
\overline{C_s}$
it follows that $\A_{|\Pi_r^{C_s}}$ embeds into $\D_{|\Pi_r^{C_s}}$.

\medskip
\noindent We proved that $\psi$ is a $\T$-general uniform interpolant
of $G$ w.r.t.\ $\Sigma_s \cup C_s$.
\QED

\begin{ex}
Consider the extension of $LI({\mathbb R})$ with uninterpreted function symbols
in $\Sigma = \{ f, g \}$. 
Let $G = (a \leq f(e)) \wedge (e \leq g(b)) \wedge (g(b) \leq a)
\wedge (f(e) \leq g(b))$.

\smallskip
\noindent We want to compute the uniform interpolant of $G$ w.r.t.\ $\{ g, a, e
\}$ ($\Sigma_s = \{ g \}, C_s = \{ a, e \}$). 
For this, we need to eliminate $\{ f, b\}$. To eliminate $f$ we use
Alg.~1.

\smallskip
\noindent {\bf Step 1:} With the definitions ${\sf Def} := \{ e_f \approx
f(e), b_g \approx g(b) \}$ we obtain, after instantiation, purification and simplification:

$(a \leq e_f) \wedge (e \leq b_g) \wedge (b_g \leq a) \wedge (e_f \leq
b_g)$

\smallskip
\noindent {\bf Step 2:} The constants corresponding to elements in
$\Sigma_s$ are $\{g_e, e, a \}$; $b$ is an argument of $g$. 
The remaining constants are: $\{ e_f\}$. 

\smallskip
\noindent {\bf Step 3:} We use quantifier elimination in $\T_0$ to eliminate $e_f$: 

\noindent $\exists e_f ((a \leq e_f) \wedge (e \leq b_g) \wedge (b_g \leq a) \wedge (e_f \leq
b_g)) \equiv (a \leq b_g) \wedge (e \leq b_g) \wedge (b_g \leq a)$ 
 
\smallskip
\noindent {\bf Step 4:} We obtain $\Gamma = (a \approx g(b) \wedge e \leq
g(b))$.

\smallskip
\noindent We need to compute $\exists b (a \approx g(b) \wedge e
\leq g(b))$ in $LI({\mathbb R}) \cup {\sf UIF}_{\{g\}}$, i.e.\ the 
$LI({\mathbb R}) \cup {\sf UIF}_{\{ g \}}$-uniform interpolant $\psi$ w.r.t.\
$C_s = \{ e, a \}$; for this
  methods proposed in 
  \cite{ghilardi-superposition,ghilardi-dag,ghilardi-combinations} can
  be used. Since by Lemma~\ref{unif-int-no-new-functions} $\psi$ 
contains only extension symbols and constants in $\{ g, e, a \}$, 
by Proposition~\ref{prop-leave-ax-forgotten-out}, $\psi$ is a 
$LI({\mathbb R}) \cup {\sf UIF}_{\{ f,g \}}$-general uniform interpolant w.r.t.\ $\{ g, e, a \}$. 
\end{ex}

\subsection{(II) Eliminating constants occurring below function symbols} 
Until now we showed that Algorithm~1 can be used to eliminate extension
functions with arity $\geq 1$ and constants which do
not occur below functions in $\Sigma_s$. 
In a second step, the remaining constants in $C_e$ need to be
eliminated. However, this is not always possible, as shown in the 
next example. 

\begin{ex} 
Consider the extension of the theory $LI([0, 1])$ with a function
symbol $f$ satisfying axioms ${\cal K} = \{ {\sf Mon}_f, \forall x
(f(x) \leq 1) \}$.
Let $G =  (a \leq b) \wedge (b \leq f(b))$. 

If we try to compute the ${\cal T}$-uniform interpolant of $G$ w.r.t.\
$\Sigma_s \cup C_s$, where $\Sigma_s = \{ f \}$, $C_s = \{ a \}$,
and ${\cal T} = LI([0, 1]) \cup {\cal K}$, we notice that no finite formula containing only
symbols in $\Pi_0$ and $a$ and $f$ has the properties of a ${\cal T}$-uniform interpolant w.r.t.\
$\Sigma_s$. 
 
Indeed, the family of all consequences $\theta$ of $G$ containing
arbitrary $\Pi_0$-symbols, but sharing with
$G$ the uninterpreted symbols $\{ a, f \}$ contains the infinite set  
$\{ a \leq f^n(1) \mid n \in {\mathbb N} \}$. 
A ${\cal T}$-uniform
interpolant $\psi$ for $G$ w.r.t.\ $\Sigma_s \cup C_s$ would only contain a finite set of 
terms, so there would exist a natural number $k$ such that 
$f^1(1), \dots, f^k(1)$ occur in $\psi$, but $f^{k+1}(1)$ does not 
occur in $\psi$. 

Note that $G \not\models_{\cal T} a = 0$ and there exist models of
$\T$ and $G$ (hence also of
$\psi$) in which $0 < f_\A^n(1) < f^{n-1}_{\cal
  A}(1) < 1$ for all $n \geq 1$. 
Let $\A$ be a model for ${\cal T}$ and $G$ (hence also of $\psi$)  
in which $a_\A > 0$ and $0 < f_\A^k(1) < \dots < f_{\cal
  A}(1) < 1$. 
We can change $\A$ to a model $\B$ 
in which the values of the terms occurring in $\psi$ have the same value, 
but the interpretation of $f$ is changed such that for every element $x < f_\B^{k-1}(1)$ we have $f_\B(x) = 0$. 
Since $f_\B^{k}(1) < f_\B^{k-1}(1)$, $f^{k+1}_\B(1) = 0$. $\B$ is still a model of $\psi$ and of
${\cal T}$, but $a > f_\B^{k+1}(1)$, so 
$\psi \not\models_{\cal T} a \leq f^{k+1}(1)$. 
\label{counterexample}
\end{ex}
In what follows, we identify situations in which constants can be
eliminated. 

\subsubsection{Extensions with
uninterpreted function symbols.} 
We first prove that the union of a theory allowing quantifier
elimination with uninterpreted function symbols in a set $\Sigma$
allows general uniform interpolation. 
\begin{thm}
If ${\cal T}_0$ is a convex, stably infinite, equality
interpolating, universal theory allowing quantifier elimination
and 
${\cal T} = {\cal T}_0 \cup {\sf UIF}_{\Sigma}$ is the extension of ${\cal
  T}_0$ with uninterpreted function symbols $\Sigma$, 
then ${\cal T}$ has general uniform interpolation. 
\label{uif}
\end{thm}
{\em Proof:} 
Let $\Pi$ be the signature of ${\cal T}$, $C$ a set of additional
constants, and let $G$ be a set of flat and linear ground $\Pi^C$-clauses. 
Let $\Sigma_s \subseteq \Sigma$ and $C_s \subseteq C$ occurring in $G$, 
and let $\Sigma_e \cup C_e$ be the symbols in $G$ which are not in
$\Sigma_s$ or $C_s$, i.e.\ which we want to
eliminate. 
We can compute a ${\cal T}$-general uniform interpolant for $G$ w.r.t.\
$\Sigma_s \cup C_s$ as follows: 
\begin{itemize}
\item We use Algorithm~1 (with ${\cal K} = \emptyset$) for eliminating 
the function symbols in $\Sigma_e$ with arity $\geq 1$ and all
constants in $C_e$ which do not occur as arguments of 
function symbols in $\Sigma_s$ and obtain a formula $\Gamma$, 
containing symbols in $\Sigma_s \cup C_s \cup C'$, where $C'$ are the
constants in $C_e$ occurring below function symbols in $\Sigma_s$. 
\item As shown in Example~\ref{ex:lir-uif}, by Theorem~\ref{thm-comb},
  $\T$ admits uniform quantifier-free interpolation. 
We use the method proposed in \cite{ghilardi-combinations}
for computing the uniform interpolant $\psi$ of  $\Gamma$ 
w.r.t.\ $C_s$ for ${\cal T}_0 \cup {\sf UIF}_{\Sigma_s}$.

\end{itemize}
We prove that the resulting formula is indeed a ${\cal T}$-general uniform 
interpolant w.r.t.\ $\Sigma_s \cup C_s$. 
Since $\Gamma$ contains only symbols in $\Pi_s^{C_s \cup C'}$, 
the uniform interpolant $\psi$ computed using the method proposed 
in \cite{ghilardi-combinations}
contains only extension symbols in $\Sigma_s$ and constants in $C_s$. 

\noindent (1) By Theorem~\ref{symb-elim-simplif}(1), $G \models_{\cal T} \Gamma$. 
The results in \cite{ghilardi-combinations} show that $\Gamma
\models_{\cal T} \psi$. 

\noindent (2) Note that since $\K = \emptyset$, the conditions in 
Theorem~\ref{symb-elim-simplif}(2) hold.  Let $\theta$ be a ground $\Pi^C$-formula having only
uninterpreted symbols
in $\Sigma_s \cup C_s$ in common with $G$, such that 
$G \models_{\cal T} \theta$. 
Since $\Sigma_s \cup C_s \subseteq \Sigma_s \cup C_s \cup C'$, 
by Theorem~\ref{symb-elim-simplif}(2) it follows that $\Gamma \models_{\T} \theta$. 
$\theta$ has in common with $\Gamma$ only constants in $C_s$ (and the
symbols in $\Sigma_s$). 
Therefore, since $\psi$ is a uniform interpolant of $\Gamma$ w.r.t.\
$C_s$ in the theory ${\cal T}_0 \cup {\sf UIF}_{\Sigma_s}$, 
we have $\psi \models_{\cal T} \theta$.  \QED

\medskip
\noindent We now present a situation in which constants can be
eliminated using a reduction to uniform interpolation in 
extensions with free function symbols of theories $\T_0$ satisfying the conditions in
Theorem~\ref{uif}.  

\subsubsection{Uniform interpolation with $\Theta$-sharing.} 
For local theory extensions $\T_0 \subseteq \T_0 \cup \K$ in which 
some of the clauses in $\K$ contain both symbols in $\K_s$ and in $\K_1$, 
or for $\Psi$-local theory extensions, where $\Psi$ is a closure
operator with the property that the set of instances ${\cal K}[\Psi(G)]$ might contain function 
symbols not occurring in $G$, 
in the proof of Theorem~\ref{symb-elim-simplif}(2) we will not be able 
to guarantee that  $\A_{|\Pi_r^{C_r}}$ embeds into  
${\cal B}_{|\Pi_r^{C_r}}$. 
A solution to this problem is to define an equivalence  
relation $\simeq$ on the uninterpreted function symbols in $\Sigma_1$, 
in such a way that if $f \simeq g$ and $f$ is shared then $g$ must
be considered to be shared as well. 
We mention two ways of defining such equivalence relations: 
\begin{enumerate}
\item Define $f \sim g$ iff there exist $c, d$ such that if $g(d) \in
  T$ then $f(c) \in \Psi(T)$ or vice versa, and 
consider the induced equivalence relation $\simeq$.
\item Define $f \sim g$ iff there exists a clause in ${\cal K}$ in
  which both $f$ and $g$ occur, and consider the induced equivalence relation $\simeq$.
\end{enumerate}
We can then define $ \Theta(\Sigma_s) = \{ f \mid \exists g \in \Sigma_s \text{ with } f
\simeq g \},$ and apply Algorithm~1 with $\Theta(\Sigma_s)$ instead of $\Sigma_s$ 
and $T = {\sf est}[\Psi(G)]$.

\begin{ex}
Consider the extension $\T := LI({\mathbb R}) \cup {\cal K}$ of $LI({\mathbb R})$ 
with monotone functions $f, g$
satisfying the axioms $\K = \{ x {\leq} g(y) \rightarrow f(x) {\leq} g(y)
\} \cup {\sf Mon}_f \cup {\sf Mon}_g$. In \cite{sofronie-lmcs08},
cf. also \cite{peuter-sofronie-cade-2023} we proved that this
is a local theory extension. 

\noindent Let $G = (a \leq f(e)) \wedge (e \leq
g(b)) \wedge (g(b) \leq a)$.
We want to compute a $\T$-general uniform interpolant of $G$ w.r.t.\ $\{ g, a, e \}$.
Since $f$ and $g$ occur together in an axiom in ${\cal K}$ we have 
$f \sim g$ and we can define, using possibility (2) above, a closure operator $\Theta$ with $\Theta(\{
g, e \}) = \{ f, g, a, e \}$. In this case $\Sigma_e = \{ b \}$, so
Algorithm 1 does not need to be used for eliminating the function symbols with arity $\geq 1$. 
We have to compute 
$\exists b ((a \leq f(e)) \wedge (e \leq
g(b)) \wedge (g(b) \leq a))$ w.r.t.\ ${\cal T}$. However,
Example~\ref{counterexample} shows that monotonicity axioms
can in general be problematic. 
\label{ex:sharing}
\end{ex}
We identify situations in which such problems can be avoided.
We focus on definitions of closure operators based on
equivalence relations on symbols defined 
according to possibility~(2) above.\footnote{In many cases the closure
  operators on terms used for possibility (1) reflect the condition in
  possibility~(2).} 
\begin{defi}
Let $\Pi_0 = (\Sigma_0, {\sf Pred})$, let $\Pi = (\Sigma_0 \cup
\Sigma, {\sf Pred})$, where $\Sigma$ is a set of
additional function symbols. Let $\K$ be a set of flat and linear
$\Pi$-clauses. For $f, g \in \Sigma$ we define $f \sim g$ if and only
if there exists a clause in ${\cal K}$ in which both $f$ and $g$ occur. 
Let $\simeq$ be the equivalence relation induced by $\sim$. 
We define a closure operator $\Theta_{\K}$ by: 
%
$\Theta_{\K}(\Sigma_s) := \{ f \mid \exists g \in \Sigma_s \text{ with } f
\simeq g \},$ for every set $\Sigma_s \subseteq \Sigma$. 
\label{theta-k}
\end{defi}
\begin{lem}
Let $\T_0 \subseteq \T_0 \cup \K$ be an extension of $\T_0$ with
symbols in a set $\Sigma$ satisfying axioms $\K$. 
 Let $G$ be a set of ground $\Pi^C$-clauses, $\Sigma_s$ and $C_s$ sets of 
function symbols and constants occurring in $G$, $\Sigma_e$ and $C_e$ the
remaining extension symbols resp.\ constants in $G$,  
$\Sigma'_s = \Theta_{\K}(\Sigma_s)$ and $\Sigma_1 =
\Theta_{\K}(\Sigma_e)$, (where $\Theta_{\K}$ is defined as
explained in
Definition~\ref{theta-k}). 
Then the set ${\cal K}$ can be written as the disjoint union 
of three sets of clauses: ${\cal K}_s$, containing only function 
symbols in $\Pi_0 \cup \Sigma'_s$,  ${\cal K}_1$, containing only 
function symbols in $\Pi_0 \cup \Sigma_1$, and ${\cal K}_i$,
containing only function symbols in $\Pi_0 \cup \Sigma_i$, 
where $\Sigma_i = \Sigma
\backslash (\Sigma'_s \cup \Sigma_1)$. 
\label{lem-disjoint}
\end{lem}
Considerations similar to those in
Theorem~\ref{inv-trans-qe} and Proposition~\ref{prop-leave-ax-forgotten-out} 
allow us to ignore (w.l.o.g.) the symbols in $\Sigma_i$ and the
clauses in $\K_i$, because those symbols are not affected by
Algorithm~1. 

\smallskip 
\noindent We identify a situation in which only constants can be
used as arguments of functions in $\Sigma'_s$, and computing a $\T$-general
uniform interpolant for $G$ w.r.t.\ $\Sigma'_s \cup C_s$ can be reduced
to computing a ${\cal T}_0 \cup
{\sf UIF}_{\Sigma'_s \cup \Sigma_1}$-uniform quantifier-free interpolant.

\begin{lem}
Let ${\cal T}_0 \subseteq {\cal T}_0 \cup {\cal K}_s \cup {\cal K}_1$
be a local extension with flat clauses, with signature $\Pi= \Pi_0 \cup \Sigma'_s \cup
\Sigma_1$, in which the extension symbols $\Sigma'_s$
occurring
in ${\cal K}_s$ and $\Sigma_1$ occurring in ${\cal K}_1$ are disjoint,
and for every clause in $\K_s \cup \K_1$, every variable occurs below
an extension symbol.
Let $T_0$ be a set of $\Pi^C$-terms, and 
$T = \{ f(t_1, \dots, t_n) \mid f \in \Sigma'_s, t_i \in T_0 \}$.
Then for every ground $\Pi^C$-formula $\theta$ whose terms starting
with function symbols in $\Sigma'_s$ are in $T$, the following are equivalent:
\begin{itemize}
\item[(1)] ${\cal T}_0 \cup {\cal K}_s \cup {\cal K}_1 \cup \theta \models \perp$. 
\item[(2)] ${\cal T}_0 \cup {\cal K}_s[T] \cup {\cal K}_1 \cup \theta \models \perp$. 
\end{itemize}
\label{instances}
\end{lem}
{\em Proof:} 
(2) $\Rightarrow$ (1) is obvious. 
Assume now that (2) does not hold, i.e.\  ${\cal T}_0 \cup {\cal K}_s[T] \cup {\cal
  K}_1 \cup \theta$ has a model $\A$.
From $\A$ we can form a partial $\Pi^C$-algebra ${\cal P}$ with the
same support as $\A$ and  in which 
the operations in $\Sigma_1$ are total and defined as in $\A$, and 
for all $p_1, \dots, p_n  \in |{\cal P}|$ and every $f \in \Sigma'_s$
we define 

\smallskip
\noindent $f_{\cal P}(p_1, \dots, p_n) = \left\{ \begin{array}{ll} 
f_{\A}(p_1, \dots, p_n) & 
\exists t_1, \dots, t_n \text{ s.t.\ } f(t_1, \dots, t_n)
\in T \text{ and } {t_i}_{\cal A} = p_i \\
\text{undefined} & \text{ otherwise}. 
\end{array} \right.$ 

\smallskip
\noindent Then ${\cal P}$ is a weak partial model of $\T_0 \cup \K_s \cup
\K_1$ with totally defined $\Sigma_0 \cup \Sigma_1$-functions. By the locality assumption, 
${\cal P}$ weakly embeds into a total model $\B$ of ${\cal T}_0 \cup 
{\cal K}_s \cup {\cal K}_1$. Since $\theta$ is a ground formula 
and all its terms are defined in ${\cal P}$ and are evaluated as in
${\cal A}$, and $\A \models \theta$ we have that $\B \models \theta$
as well, so $\B$ is a model of ${\cal T}_0 \cup {\cal K}_s \cup {\cal
  K}_1 \cup \theta$, i.e.\ (1) does not hold. \QED

\medskip
\noindent For the equivalence in Lemma~\ref{instances} to hold for
all ground $\Pi^C$-formulae $\theta$, 
$T_0$ has to be the set of all $\Pi^C$-terms, so $T$ and therefore also ${\cal
  K}_s[T]$ will in general be infinite. In what follows we identify a 
simple case in which the set $T$ of all terms starting with a function
symbol in $\Sigma_s'$ is finite, hence if $\K_s$ is finite 
then ${\cal K}_s[T]$ is finite as well. 

\smallskip
\noindent We consider $\T$-general uniform interpolation problems for sets $G$ of ground clauses 
w.r.t.\ extension functions $\Sigma_s$ and constants $C_s$ occurring in
$G$, and consider $\Sigma_s' = \Theta_{\K}(\Sigma_s)$, 
(cf.\ Definition~\ref{theta-k}), where $\T = \T_0 \cup \K$ is a theory extension 
over a many-sorted signature with the following properties: 

\begin{description}
\item[(A1)] ${\cal T} = {\cal T}_0 {\cup} \K$ is a theory over a signature $\Pi
  =(S, \Sigma_0 {\cup} \Sigma'_s {\cup} \Sigma_1, {\sf Pred})$ with set of sorts
  $S = S_1 \cup S_2$, where $S_1 \cap S_2 = \emptyset$ (an extension
  of a theory $\T_0$ with signature $\Pi_0$ with a finite set of function symbols 
  $\Sigma = \Sigma'_s {\cup} \Sigma_1$). 
\item[(A2)] $\K$ is a finite set of flat and linear $\Pi$-clauses,  
$\K = \K_s {\cup} \K_1$,
where $\K_s$ is a set of $\Pi_0 {\cup} \Sigma'_s$-clauses, $\K_1$ 
a set of $\Pi_0 {\cup} \Sigma_1$-clauses, for every clause in $\K_s {\cup} \K_1$, every variable occurs below
an extension symbol, and the following hold: 
\begin{itemize}
\item[(a)] For every  $f$ in $\Sigma_0 {\cup} \Sigma'_s {\cup}
  \Sigma_1$ with arity $\geq 1$, the output sort of $f$ is in $S_2$. 
\item[(b)] For every $f \in \Sigma'_s$ with arity $a(f) = s_1,
  {\dots}, s_n {\rightarrow} s$ we have $s_1, {\dots}, s_n {\in} S_1$.
\item[(c)] The constants of sort in $S_1$ are in $\Sigma'_s \cup
  C$.
\end{itemize}
\end{description}

\begin{prop}
Let $\Pi$ be a many-sorted signature, 
$G$ be a finite set of flat and linear ground $\Pi^C$-clauses, $C_s \subseteq C$ a set of
constants occurring in $G$, and 
${\cal T}_0 \subseteq {\cal T} := {\cal T}_0 \cup {\cal K}_s \cup {\cal K}_1$
be a local theory extension with signature $\Pi= \Pi_0 \cup \Sigma'_s \cup
\Sigma_1$, which satisfies conditions 
${\bf (A1)}$ and ${\bf (A2)}$. 
Let 

\smallskip
$\begin{array}{ll} 
T := \{ f(c_1, \dots, c_n) \mid & f \in \Sigma'_s, a(f) = s_1 \dots s_n
\rightarrow s, \text{ s.t.\  for } 1 \leq i
\leq n\\
& c_i \text{ is a constant of sort } s_i \text{  in  } \Sigma'_s \cup C \}.
\end{array}$ 

\noindent Let $\psi$ be a ground $(\Pi_0 \cup \Sigma'_s)^{C_s}$-formula. 
If $\psi$ is a $\T_0 \cup {\sf UIF}_{\Sigma_s} \cup \K_1$-general uniform interpolant of
  $G \wedge \K_s[T]$ w.r.t.\
  $\Sigma'_s \cup C_s$, then $\psi$ is a ${\cal T}_0 \cup \K_s \cup \K_1$-general uniform interpolant of $G$ w.r.t.\
  $\Sigma'_s \cup C_s$. 
\label{prop-tame}
\end{prop}
{\em Proof:} 
First, note that due to the restrictions on the signature of $\T$, all
ground terms starting with a function symbol $f \in \Sigma'_s$ are in
$T$. Therefore, for every ground $\Pi^C$-formula $\theta$, the 
terms starting with a function symbol $f \in \Sigma'_s$ occurring in 
$\theta$ are in $T$. Moreover, $T$ and $\K_s[T]$ are finite.

Assume that $\psi$ is a $\T_0 \cup {\sf UIF}_{\Sigma_s} \cup \K_1$-general uniform interpolant of
$G \wedge \K_s[T]$ w.r.t.\
  $\Sigma'_s \cup C_s$. Then it is a $(\Pi_0 \cup
  \Sigma'_s)^{C_s}$-formula and (1) $G \wedge \K_s[T] \models_{\T_0 \cup {\sf UIF}_{\Sigma_s} \cup \K_1} \psi$, and (2) for every
  $\Pi^C$-formula $\theta$ having only uninterpreted symbols in
  $\Sigma'_s \cup C_s$ in common
  with $G \wedge \K_s[T]$, if $G \wedge \K_s[T] \models_{\T_0 \cup {\sf UIF}_{\Sigma_s} \cup \K_1}
  \theta$ then $\psi  \models_{\T_0 \cup {\sf UIF}_{\Sigma_s} \cup \K_1} \theta$. 

\smallskip
\noindent (1) We show that $G \models_{\T} \psi$. 
Indeed, as $G \wedge \K_s[T] \models_{\T_0 \cup {\sf UIF}_{\Sigma_s} \cup \K_1} \psi$ we know
that $\T_0 \cup \K_s[T] \cup
\K_1 \cup G \cup \neg \psi$ is unsatisfiable, hence (since all
subterms of $\psi$ and $G$ starting with a function symbol in $\Sigma'_s$ are
in $T$), by Lemma~\ref{instances} we know that 
$\T_0 \cup \K_s \cup
\K_1 \cup G \cup \neg \psi \models \perp$, i.e.\ $G \models_{\T} \psi$.

\smallskip
\noindent (2)  Let $\theta$ be a ground formula which has in common
with $G$ only the symbols in $\Sigma'_s \cup C_s$. 
Assume that $G \models_{\T}\theta$. Then $\T_0 \cup \K_s \cup \K_1 \cup G \cup \neg \theta \models
\perp$, so (since all subterms of $G$ and $\theta$ starting with function
symbols in $\Sigma'_s$ are in $T$), by Lemma~\ref{instances},  
$\T_0 \cup \K_s[T] \cup \K_1 \cup G \cup \neg \theta \models
\perp$, so $G \wedge \K_s[T] \models_{\T_0 \cup {\sf UIF}_{\Sigma_s} \cup \K_1} \theta$. 
Since $G \wedge \K_s[T]$ and $\theta$ share
only symbols in $\Sigma'_s \cup C_s$, and $\psi$ is a $\T_0 \cup {\sf UIF}_{\Sigma_s} \cup \K_1$-general
uniform interpolant for $G \wedge \K_s[T]$, it follows that $\psi 
\models_{\T_0 \cup {\sf UIF}_{\Sigma_s} \cup \K_1} \theta$, so $\T_0 \cup \K_1 \cup \psi \cup \neg \theta
\models \perp$, hence also 
$\T_0 \cup \K_s[T] \cup \K_1 \cup \psi \cup \neg \theta \models
\perp$, i.e.\ (again by Lemma~\ref{instances}) 
$\T_0 \cup \K_s \cup \K_1 \cup \psi \cup \neg \theta \models \perp$.
Therefore, $\psi \models_{\T} \theta$. \QED

\begin{thm}
Let ${\cal T}_0$ be a convex, stably infinite, equality
interpolating, universal many-sorted theory with signature $\Pi_0 =
(S, \Sigma_0, {\sf Pred})$ allowing quantifier elimination. Let $\Pi =
(S, \Sigma_0 \cup \Sigma, {\sf Pred})$, 
$G$ be a set of flat and linear ground $\Pi$-clauses with
additional constants $C$, 
$\Sigma_s \subseteq \Sigma$, $C_s \subseteq C$ be 
extension functions resp.\ constants occurring in $G$ and 
$\Sigma'_s = \Theta_{\K}(\Sigma_s)$ 
Assume that ${\cal T}_0 \subseteq {\cal T} := {\cal T}_0 \cup {\cal K}$
is a local theory extension with signature $\Pi$
satisfying conditions ${\bf (A1)}$ and ${\bf (A2)}$. 

Then a $\T$-general uniform
interpolant of $G$ w.r.t.\ $\Sigma'_s \cup C_s$ can be computed 
by computing a $\T_0 \cup {\sf UIF}_{\Sigma'_s} \cup \K_1$-general uniform interpolant
of $G \wedge \K_s[T]$ w.r.t.\ $\Sigma'_s \cup C_s$, where
$
T = \{ f(c_1, \dots, c_n) \mid  f \in \Sigma'_s, a(f) = s_1 \dots s_n
\rightarrow s, \text{ s.t.\  for } 1 \leq i
\leq n,  c_i \text{ is a constant of sort } s_i \text{  in  }
\Sigma'_s \cup C \}$.
This can be achieved by 
\begin{itemize}
\item[(i)] computing a $\Pi_s^{\overline{C_s}}$-formula $\Gamma$, by  applying 
Algorithm~1 to the ground formula $G
\wedge \K_s[T]$, theory $\T_0 \cup {\sf UIF}_{\Sigma'_s} \cup \K_1$,
and set $C_s$ of constants, with set of terms for the
instantiation ${\sf
  est}(G \wedge \K_s[T])$ (where $C_s
\subseteq \overline{C_s}$), and then 
\item[(ii)] computing a $\T_0 \cup {\sf  UIF}_{\Sigma'_s \cup
    \Sigma_1}$-uniform interpolant of $\Gamma$
w.r.t.\ $C_s$  (e.g.\ with
the method in \cite{ghilardi-combinations}).
\end{itemize}
\label{thm:tame}
\end{thm}
{\em Proof:} 
Let $\Pi_s = (S, \Pi_0 \cup \Sigma'_s, {\sf Pred})$ and let $\Gamma$ be the
$\Pi_s^{{\overline C}_s}$-formula (where $C_s \subseteq
\overline{C_s}$) obtained
using Algorithm~1 from $G \wedge \K_s[T]$, and the theory $\T_0 \cup
{\sf UIF}_{\Sigma'_s} \cup \K_1$ and $C_s$, with set of terms for the
instantiation ${\sf
  est}(G \wedge \K_s[T])$. Then, by Theorem~\ref{symb-elim-simplif}(2), $\Gamma$ is a 
$\T_0 \cup {\sf UIF}_{\Sigma'_s} \cup \K_1$-general uniform interpolant of $G \wedge \K_s[T]$ w.r.t.\ $\Sigma'_s \cup
{\overline C}_s$. 
Let $\psi$ be a $\T_0 \cup
  {\sf UIF}_{\Sigma'_s \cup \Sigma_1}$-uniform interpolant for $\Gamma$ w.r.t.\ $C_s$.
Then, by Proposition~\ref{prop-leave-ax-forgotten-out}, 
$\psi$ is a $\T_0 \cup {\sf UIF}_{\Sigma'_s} \cup \K_1$-general
uniform interpolant of  $G \wedge \K_s[T]$
w.r.t.\ $\Sigma'_s \cup C_s$. 
From Proposition~\ref{prop-tame} it therefore follows that 
$\psi$ is a $\T$-general uniform interpolant of $G$
w.r.t. $\Sigma'_s \cup C_s$.  \QED

\begin{ex}
Let $\T_0$ be the 2-sorted disjoint combination of $LI({\mathbb R})$ (sort
${\sf n}$) and the (model completion of the) pure theory of equality
(sort ${\sf p}$). 
Let $\Sigma = \{ f, g, h \}$, where $a(f) = {\sf n} \rightarrow {\sf
  n}$ and $a(g) = a(h) = {\sf p} \rightarrow {\sf n}$. 

\noindent Let ${\cal T}_0 \subseteq \T_0 \cup \K_s \cup \K_1$, where $\K_s =
\forall x (g(x) \leq h(x))$ and $\K_1 = {\sf Mon}_f$, 

\smallskip
\noindent $G = (c \leq g(p)) \wedge (h(p) \leq b) \wedge (p \approx q) \wedge (b \leq
g(r)) \wedge (r \approx q) \wedge (f(b) \leq c) \wedge (f(b) \approx b)$, 

\smallskip
\noindent and $\Sigma_s =
\{ g, h \}$, $C_s = \{ q, b \}$, where $a(q) = {\sf p}$ and $a(b) = {\sf
  n}$.

\smallskip
\noindent $T = \{ g(p), g(q), g(r), h(p), h(q), h(r) \}$. 
We apply Algorithm~1 to $\T_0 \cup {\sf UIF}_{g, h} \cup \K_1$, $G
\wedge \K_s[T]$ and ${\sf
  est}(G \wedge \K_s[T])$. We need to eliminate $f$ and $\{ p, r, c \}$.

\smallskip
\noindent {\bf Step 1:} After instantiation and introducing definitions 
${\sf Def} = \{ p_g \approx g(p), q_g \approx g(q), r_g \approx g(r),
p_h \approx h(p), 
q_h \approx h(q), r_h \approx h(r), b_f \approx f(b) \}$ we obtain:

$(c \leq p_g) \wedge (p_h \leq b) \wedge (p \approx q) \wedge (b \leq
r_g) \wedge (r \approx q) \wedge (b_f \leq c) \wedge (b_f \approx b) \wedge$ 

$(p_g \leq p_h) \wedge (q_g \leq q_h) \wedge (r_g \leq r_h) 
\wedge (p_g \approx q_g) \wedge (p_h \approx q_h) \wedge (r_g \approx q_g) 
\wedge (r_h \approx q_h).$ 

\smallskip
\noindent {\bf Step 2:} We classify the constants. The constants introduced for
the parameters and their arguments are $\{ p_g, q_g, r_g, p_h, q_h,
r_h, p, q, r \}$. 

\smallskip
\noindent {\bf Step 3:} We eliminate $\{ b_f, c \}$ and obtain: 

$(b \leq p_g) \wedge (p_h \leq b) \wedge (p \approx q) \wedge (b \leq
r_g) \wedge (r \approx q) \wedge$ 

$(p_g \leq p_h) \wedge (q_g \leq q_h) \wedge (r_g \leq r_h) \wedge
(p_g \approx q_g) \wedge (p_h \approx q_h) \wedge (r_g \approx q_g)
\wedge (r_h \approx q_h).$ 

\smallskip
\noindent {\bf Step 4:} We replace the new constants with the terms
they denote and ignore formulae already implied by $\T_0 \cup \K_s$
and obtain:

$(b \leq g(p)) \wedge (h(p) \leq b) \wedge (p \approx q) \wedge (b \leq
g(r)) \wedge (r \approx q).$ 

\smallskip
\noindent  We can compute $\exists p, r (b \leq g(p) \wedge h(p) \leq b \wedge p \approx q \wedge b \leq
g(r) \wedge r \approx q)$. 

\noindent It can be seen that it is equivalent to 
$(b \leq g(q)) \wedge (h(q) \leq  b) \wedge (b \leq g(q))$.
\end{ex}

\section{Definable functions} 
\label{definitions}

If we want to use Proposition~\ref{prop-leave-ax-forgotten-out} for computing a
general uniform interpolant, 
if $\K_s \neq \emptyset$  it might be problematic to compute a $\T_0 \cup
\K_s$-uniform interpolant after applying Alg.~1.
We show that if $\Sigma_s = \Sigma_f \cup
\Sigma^d_s$, $\K_s = {\sf UIF}_{\Sigma_f} \cup \K^d_s$, and all
symbols in $\Sigma^d_s$ are definable w.r.t.\ $\T_0 \cup \K_s$ over
the signature $\Pi_0 \cup \Sigma_f$,  computing $\T_0 \cup \K_s$-uniform
interpolants can be reduced to computing $\T_0 \cup {\sf
  UIF}_{\Sigma_f}$-interpolants.

\begin{defi}
Let ${\cal T}$ be a theory with signature $\Pi = (\Sigma, {\sf Pred})$, 
and $f \in \Sigma$ with arity $n$.
We say that  $f$ is {\em explicitly definable} w.r.t.\ $\T$ over the signature $\Pi_1
\subseteq \Pi$ iff there exists 
a $\Pi_1$-formula $F_f$ with free variables $x_1, \dots, x_n, y$ 
such that 

\smallskip
${\cal T} \models \forall x_1, \dots, x_n, y ((f(x_1, \dots,
x_n) \approx y) \leftrightarrow F_f(x_1, \dots, x_n, y)).$

\smallskip
\noindent We say that $f$ is {\em implicitly definable} w.r.t.\ $\T$ over the signature
$\Pi_1$ if whenever ${\cal T}'$ is the set of formulae obtained 
from $\T$ by renaming the function and predicate symbols which are not 
in $\Pi_1$ (including $f$) to primed versions, 

\smallskip
${\cal T} \cup {\cal T}' \models 
\forall  x_1, \dots, x_n (f(x_1, \dots,
x_n) \approx f'(x_1, \dots, x_n)).$ 
\end{defi} 
\begin{prop}
Let $T_0$ be a $\forall\exists$ theory with signature $\Pi_0$, 
in which ground satisfiability
is decidable.  Let $\Sigma$ be a set of extension functions and let 
$\Sigma_1 \subseteq \Sigma$.  Let  
$\T_0 \subseteq \T_0 \cup {\sf UIF}_{\Sigma_1} \subseteq \T = \T_0 \cup
{\sf UIF}_{\Sigma_1} \cup \K$
be a chain of local theory
extensions, where 
$\K$ is a set of flat and linear clauses, such that in every clause
every variable occurs below an extension function in $\Sigma
\backslash \Sigma_1$.
Then testing whether a function $f \in \Sigma \backslash \Sigma_1$
is implicitly definable w.r.t.\ $\T$ over $\Pi_1 = \Pi_0 \cup \Sigma_1$  
is decidable.
\end{prop}
{\em Proof:} 
Let $\K'$ be obtained by replacing every function 
$f \in (\Sigma \backslash \Sigma_1)$ 
by a primed version. If $\T_0 \cup {\sf UIF}_{\Sigma_1} \subseteq \T_0
\cup {\sf UIF}_{\Sigma_1} \cup \K$ is local then
also $\T_0 \cup {\sf UIF}_{\Sigma_1} \subseteq \T_0 \cup {\sf
  UIF}_{\Sigma_1} \cup \K'$ is local. By Theorem~\ref{thm:combine} 
(cf.\ \cite{sofronie-ihlemann-10}) we know that 
$\T_0 \cup {\sf UIF}_{\Sigma_1} \subseteq \T_0 \cup {\sf
  UIF}_{\Sigma_1} \cup \K \cup \K'$ is local. 
Then the following are equivalent
\begin{itemize}
\item The function 
$f$ is implicitly definable in $\Pi_1$. 
\item $\T_0 \cup {\sf  UIF}_{\Sigma_1} \cup \K \cup \K' \models \forall
x (f(x_1,\dots,x_n) \approx f'(x_1,\dots,x_n))$. 
\item  $\T_0 \cup \K \cup \K' \cup f'(a_1,\dots,a_n) \not\approx f(a_1,\dots,a_n)
\models \perp$, where $a_1,\dots,a_n$ are new Skolem constants. 
\end{itemize}
We can use the method for hierarchical reasoning in local theory
extensions to effectively check whether this is the case. \QED

\begin{thm}
Let $\T_0 \subseteq \T_0 \cup {\sf UIF}_{\Sigma_f} \subseteq \T := \T_0 \cup {\sf UIF}_{\Sigma_f} \cup 
\K$ be a chain of local theory extensions, such that the signature of
$\T_0$ is $\Pi_0$, the signature of $\T$ is $\Pi = \Pi_0 \cup \Sigma$,
where $\Sigma = \Sigma_f
\cup \Sigma_d$, and $\K$ is a set of $\Pi$-clauses such that
for every clause in $\K$, every variable occurs below
an extension symbol in $\Sigma_d$.
If $f \in \Sigma_d$ is implicitly definable w.r.t.\ $\T$ over $\Pi_0 \cup \Sigma_f$
then we can use Algorithm~1 to eliminate the function symbols in
$\Sigma_d$, followed by uniform 
interpolant computation w.r.t.\ $\T_0 \cup {\sf UIF}_{\Sigma_f}$ 
to find a $\Pi_0 \cup \Sigma_f$-formula $F_f$ 
such that 
${\cal T} \models \forall {\overline x}, y (
(f({\overline x}) \approx y) \leftrightarrow F_f({\overline x}, y))$. 
\label{thm:definability}
\end{thm}
{\em Proof:} 
The proof uses the fact that if 
$\T_0 \subseteq \T_0 \cup {\sf UIF}_{\Sigma_f} \subseteq \T := \T_0
\cup {\sf UIF}_{\Sigma_f} \cup \K$ is a chain of local theory
extensions, then $\T_0 \subseteq \T_0 \cup  {\sf UIF}_{\Sigma_f}
\cup \K$ is a $\Psi$-local theory extension, for the closure
operator $\Psi$ with the property that for every set $T$ of ground terms and 
every if $f(t_1, \dots, t_n) \in T$ with $f \in \Sigma_d$, then 
for every other function symbol $g \in \Sigma_f \cup \Sigma_d$ with
arity $k$, and all $t'_1, \dots, t'_k \in \{ t_1, \dots, t_n \}$ we
have $g(t'_1, \dots, t'_k) \in \Psi(T)$. Since
$\Sigma_1 \cup \Sigma_d$ are finite, it is clear that if $T$ is
a finite set of flat terms then $\Psi(T)$ is finite as well.

For $G = \{f(a_1, \dots, a_n) \approx b \}$, $\Psi({\sf est}(G)) = \{
g(a_{i_1}, \dots, a_{i_k}) \mid g \in \Sigma_f \cup \Sigma_d \text{
  has arity } k, a_{i_j} \in \{ a_1, \dots, a_n \} \}$. 

\smallskip
We can proceed as follows: 
\begin{enumerate}
\item We start with $\T_0 \cup {\sf UIF}_{\Sigma_f} \cup \K \cup f(a_1, \dots, a_n) \approx b$,  
and $\Sigma_s := \Sigma_f$, $C_s := \{ a_1, \dots, a_n, b \}$. 
\item We use Algorithm~1 to eliminate all function symbols in 
$\Sigma_d := \Sigma \backslash \Sigma_f$, for $\T_0 \cup {\sf UIF}_{\Sigma_f} \cup
\K$, $G = f(a_1, \dots, a_n) \approx b$ and for $T = {\sf
  est}(\Psi(G))$.  
Algorithm~1 yields a $(\Pi_0 \cup \Sigma_f)^C$-formula
$\Gamma({\overline a}, b, {\overline c})$. 
\item We use $\T_0 \cup {\sf
  UIF}_{\Sigma_f}$-uniform interpolation w.r.t.\ $\{ a_1, \dots, a_n, b \}$ 
to eliminate the constants ${\overline c}$ which are not in $\{ a_1, \dots, a_n, b \}$. 
\end{enumerate}
We prove that if $f$ is implicitly definable, 
the formula $F_f({\overline a}, b)$ obtained  this way is the required
explicit definition of $f$.
For this, we analyze the way Algorithm~1 can be applied. 
We compute $\K[\Psi[G]]$, then purify 
$\K[\Psi[G]] \wedge f(a_1, \dots, a_n) \approx b$ by
introducing new constants with their definitions ${\sf Def}$, e.g.\ 
$c_g = g(a_{i_1}, \dots, a_{i_k}) \in {\sf Def}$. 
Among these constants only $a_1, \dots, a_n, b$ are in $C_s = \{ a_1, \dots, a_n, b \}$. 
Let ${\overline c}$ be the other constants which occur below function
symbols in $\Sigma_f$. 
Let $\Gamma({\overline a}, b, {\overline c})$ be the formula
obtained by applying Algorithm~1. The constants ${\overline c}$ have to be eliminated. 

\medskip
\noindent {\bf Claim 1:} $\T_0 \cup \K \models \forall
{\overline x}, y (f({\overline x}) \approx y \rightarrow F_f({\overline  x}, y))$. \\
{\em Proof:} By Theorem~\ref{symb-elim-simplif}(1), 
$(f(a_1, \dots, a_n) \approx b) \models_{\T_0 \cup
  \K}  \Gamma(a_1, \dots, a_n, b, {\overline c})$.
Since $F_f({\overline a}, b)$ is the $\T_0 \cup {\sf
  UIF}_{\Sigma_f}$-uniform interpolant of
$\Gamma(a_1, \dots, a_n, b, {\overline c})$ w.r.t.\, $C_s = \{ a_1,
\dots, a_n, b \}$, 
we know that $\Gamma(a_1, \dots, a_n, b, {\overline c})
\models_{\T_0 \cup {\sf UIF}_{\Sigma_f}} F_f({\overline a}, b)$,
therefore we also have 
$\Gamma(a_1, \dots, a_n, b, {\overline c})
\models_{\T_0 \cup {\sf UIF}_{\Sigma_f} \cup \K} F_f({\overline a}, b)$.
From this it follows that  $(f(a_1, \dots, a_n) \approx b) \models_{\T_0 \cup {\sf
    UIF}_{\Sigma_f} \cup  \K} F_f({\overline a}, b)$, so 
$\T_0 \cup \K \models \forall
{\overline x}, y (f({\overline x}) \approx y \rightarrow F_f({\overline
  x}, y))$. 

In particular, since $f$ is a total function, 
it follows that for every $\Pi^C$-structure $\A$ and 
all elements $m_1, \dots, m_n \in |\A|$ 
there exists at least one element $n \in |\A|$ 
such that $F_f({\overline m}, n)$ holds, namely $n = f_\A(m_1, \dots,
m_n)$ -- but there can exist more such elements.

\medskip
\noindent {\bf Claim 2:} If $f$ is implicitly definable,  
$\T_0 {\cup} \K \models \forall {\overline x}, y (F_f({\overline
  x}, y) {\rightarrow}  f({\overline x}) {\approx} y)$.  \\
{\em Proof:} We show that $F_f(a_1, \dots, a_n, b) \models_{\T_0 \cup \K}
f(a_1, \dots, a_n) \approx b$.  
Let ${\cal A}$ be a $\Pi^C$-structure which is a model of 
$\T = \T_0 \cup {\sf UIF}_{\Sigma_f} \cup \K$ with 
${\cal A} \models F_f(a_1, \dots, a_n, b)$, and let ${a_1}_{\cal
  A} = m_1, \dots, {a_n}_{\A} = m_n$, $b_{\cal A} = k$. 

\smallskip
\noindent {\em Step 1:} By Theorem~\ref{ghilardi-emb},
there exists another $\Pi^C$-structure $\B$ which is a model of $\T$ such that ${\cal A}_{|\Pi}$ 
embeds into ${\cal B}_{|\Pi}$ and ${\cal B} \models \Gamma({\overline a}, b, {\overline c})$, such that 
the interpretations of $a_1, \dots, a_n, b$ are the same in
${\cal A}$ and ${\cal B}^{\overline c}$, but the constants in ${\overline c}$ might
have different interpretations. 

\smallskip
\noindent {\em Step 2:} We can use the arguments in Theorem~\ref{symb-elim-simplif}(2), this
time for $T = {\sf est}(\Psi(G))$ instead of ${\sf est}(G)$ 
to show that 
$\Gamma({\overline a}, b, {\overline c})$ is a $\T$-general uniform interpolant of
$f({\overline a}) \approx b$ w.r.t.\ $\Sigma_s \cup \{ a_1, \dots, a_n, b \} \cup
{\overline c}$. 
We show how the proof changes in this case (the $\Psi$-locality
condition allows us to lift the restriction that $\K$ cannot
contain symbols $\Sigma_f$ necessary in the proof of Theorem~\ref{symb-elim-simplif}(2)).
We can, as in the proof of Theorem~\ref{symb-elim-simplif}(2), purify
$\Gamma$ by introducing new constants $C'$ for the $\Sigma_f$-terms, 
and obtain a ground formula 
$\Gamma_1 \cup {\sf Def}_f$, where ${\sf Def}_f$ are definitions for
the constants $C'$ renaming $\Sigma_f$-terms in $\Gamma$.
We know that $\Gamma_1 \equiv_{\T_0} \exists {\overline d} (\K_0 \cup
G_0 \cup {\sf Con}_0)$. Let $\B^{C'}$ be
the extension of $\B$ with the new constants $C'$ defined
as described in the definitions ${\sf Def}_f$. 
Then $\B^{C'}_{|\Pi^{C'}_0} \models \exists {\overline d} (\K_0 \cup
G_0 \cup {\sf Con}_0)$. This shows that there are values ${\overline
  a}$ for the constants in ${\overline d}$ in $|\B^{C'}|$ 
(perhaps different from the interpretations of these constants in
$\B^{C'}$) for which $({\cal K}_0 \cup
G_0 \cup {\sf Con}_0)$ holds. 
Let $C''$ be $C'$ to which we add the constants ${\overline d}$
determined in Step 2 of Algorithm~1. 
Let ${\overline {\cal B}}$ be a $\Pi^{C \cup C''}$-structure with the
same support as ${\cal B}$, 
which agrees with $\B^{C'}$ on $\Pi_0, \Sigma_f$, 
$a_1, \dots, a_n, b$ and $C'$,  and such that the interpretation of 
the constants in ${\overline d}$ is ${\overline a}$. 
Then  ${\overline {\cal B}} \models {\cal K}_0 \cup G_0 \cup {\sf Con}_0$. 
Note that the constants in ${\overline d}$ do not occur among the
constants in $C' \cup C_s$ or as arguments of function symbols in
$\Sigma_f$, so 
${\overline {\cal B}}$  is still a model of ${\sf Def}_f$. 

We construct a partial
$\Pi^C$-algebra ${\cal P}$, with the same support as ${\overline \B}$
where:
\begin{itemize}
\item all symbols in $\Pi_0$ and $\Sigma_f$ 
and all constants in  ${\overline d}$ are defined as in ${\overline {\cal B}}$;
\item partial functions in $\Sigma_d$ are defined according to the definitions in
${\sf Def}$. 
\end{itemize}
The definition of partial functions in $\Sigma_f$, as reconstructed from
the definitions, coincides with the interpretation of those function
symbols in $\B$, 
because the constants in ${\overline d}$ do not occur below
function symbols in $\Sigma_f$.
The partial structure ${\cal P}$ satisfies 
${\cal K}[\Psi(G)] \cup G$: 
\begin{itemize}
\item The functions
in $\Sigma_0 \cup \Sigma_f$ 
are totally defined in ${\cal P}$, and they are
defined as in $\B$, which is a model of $\T_0 \cup {\sf UIF}_{\Sigma_1}$, so ${\cal P}$ is a
model of $\T_0 \cup {\sf UIF}_{\Sigma_1}$.
\item From the definition of $\Psi(G)$, Only the $\Sigma_d$-terms in $\Psi(G)$ are
defined in ${\cal P}$. 
\end{itemize}
We show that ${\cal P}$ weakly satisfies $\K$: 
Let $C \in \K$, and let $\beta : X \rightarrow |{\cal P}|$.

\noindent {\em Case 1:} If some term in $C$ is undefined under
$\beta$, 
$({\cal P}, \beta) \models_w C$. 

\noindent {\em Case 2:} Assume that all terms in $C$ are defined under $\beta$.
We know that every variable $x$ of $C$ occurs below a term $f(x_1, \dots,
x_k)$ with $f \in \Sigma_d$. By the way the function symbols in
$\Sigma_d$ are defined in $|{\cal P}|$, 
$f_{\cal P}(\beta(x_1), \dots, \beta(x_n))$ is defined in ${\cal P}$
and equal to $p$ 
iff there exist terms $t_1, \dots, t_k \{ a_1, \dots, a_n \}$ such
that $f(t_1, \dots, t_n) \in \Psi(G)$, and
there exists a definition $c_f = f(t_1, \dots, t_k) \in {\sf Def}$
such that ${t_i}_{\B} = \beta(x_i)$ and 
${c_f}_{\cal P} = p$. By the definition of $\Psi$, if $f_i(t^i_1, \dots,
t^i_{k_i})$ are all the terms in  $\Psi(G)$ corresponding to the terms 
$f_i(x^i_1, \dots, x^i_k)$ with $f_i \in \Sigma_d$ occurring in $C$, 
then $g(t'_1, \dots, t'_k) \in \Psi(G)$ for all $g \in \Sigma_f \cup
\Sigma_d$ and  $t'_1, \dots t'_k \in \bigcup_i \{ t^i_1, \dots t^i_{k_i})$. 

Since $\K$ is flat and linear w.r.t.\ the
$\Sigma^d_s$-symbols, we can find a substitution $\sigma : X
\rightarrow \{ a_1, \dots, a_n \}$ such that $\beta(C) =
\beta(C\sigma)$. Since $C\sigma \in \K[\Psi(G)]$, it follows that also
$C$ holds in ${\cal P}$. 

\smallskip
\noindent By the $\Psi$-locality of the extension ${\cal T}_0 \subseteq  {\cal
  T}_0 \cup {\sf UIF}_{\Sigma_1} \cup {\cal K}$, ${\cal P}$ weakly
embeds into a total model 
$\C$ of ${\cal T}_0 \cup {\sf UIF}_{\Sigma_1} \cup {\cal K}$
which is a model of $G = f(a_1, \dots, a_n) \approx b$. 

In addition, the reduct of ${\cal B}$ to the
signature $\Pi \cup \Sigma_1 \cup \{ a_1, \dots, a_n, b \} \cup
{\overline c}$ 
of symbols which are not eliminated 
embeds into the reduct of ${\cal C}$ to this signature.

\medskip
\noindent {\em Step 3:} The interpretations of the functions in $\Sigma_1$, as well as the 
interpretations of $a_1, \dots, a_n, b$ and ${\overline c}$ 
are therefore the same in ${\cal C}$ and ${\cal B}$, 
so we have $f_{\cal C}(m_1, \dots, m_n) = k$. 
Since $f$ is eliminated, it could be that in ${\cal C}$ we in fact
defined a new interpretation $f_{{\cal C}}$ of the function symbol
$f$, possibly different from the interpretation of $f$ in ${\cal A}$.
We know that the reduct of $\A$ to $\Pi_1^{C_s}$ embeds into 
the reduct of ${\cal C}$ to $\Pi_1^{C_s}$. 
Let $\Sigma_e = \{ g \mid g \in \Sigma \backslash \Sigma_1 \}$ and 
$\Sigma'_e = \{ g' \mid g \in \Sigma \backslash \Sigma_1 \}$, and 
let $\K'$ be obtained from $\K$ by renaming all symbols in $\Sigma_e$ 
to their primed versions. 

Let ${\cal P}$ be a partial structure with the same universe as ${\cal C}$, in
which all symbols in $\Pi_0 \cup \Sigma_1$ and the constants $\{ a_1,
\dots, a_n, b \}$ are defined as in ${\cal C}$, and 
in which, for every $g \in \Sigma_e$: 
\begin{itemize}
\item for every $v_1, \dots, v_p \in |\A|$ we
have $g_{\cal P}(v_1, \dots, v_p) = g_{\cal A}(v_1, \dots, v_p)$, and
$g_{\cal P}$ is undefined otherwise, and 
\item for every $w_1, \dots, w_p \in |{\cal C}|$, $g'_{\cal D}(w_1, \dots, w_p) =
g_{{\cal C}}(w_1, \dots, w_p)$. 
\end{itemize}
${\cal P}$ is a weak partial model
of $\T_0 \cup {\sf UIF}_{\Sigma_s} \cup \K \cup \K'$ in which all
functions in $\Pi_0 \cup \Sigma_1$ are defined. 
Since the extension $\T_0 \cup {\sf UIF}_{\Sigma_1} \subseteq \T_0
\cup {\sf UIF}_{\Sigma_1} \cup \K \cup \K'$ is local, ${\cal P}$ weakly
embeds into a total model ${\cal D}$ 
of $\T_0 \cup {\sf UIF}_{\Sigma_1} \cup \K \cup \K'$, 
in which the constants $a_1, \dots, a_n, b$ are interpreted 
as in ${\cal C}$ and in ${\cal A}$.

\medskip
\noindent Since $f$ is implicitly definable, $f_{\cal D}(m_1, \dots, m_n) =
f'_{\cal D}(m_1, \dots, m_n)$, hence we have $f_{\cal C}(m_1,\dots, m_n) =
f_{\cal A}(m_1, \dots, m_n)$. Since we know that $\C \models f(a_1, \dots, a_n) \approx
b$, we have $f_{\A}(m_1, \dots, m_n) = f_{\C}(m_1, \dots, m_n) = k$. 
Thus, ${\cal A} \models f(a_1, \dots, a_n) \approx b$. \QED

\begin{thm}
Let ${\cal T}_0 \subseteq \T_0 \cup {\sf UIF}_{\Sigma_1} \subseteq
{\cal T} := {\cal T}_0 \cup {\sf UIF}_{\Sigma_f} \cup {\cal K}_s
\subseteq \T_0 \cup {\sf UIF}_{\Sigma_f} \cup \K_s \cup \K_1$ be a
chain of local theory extensions with extension functions in a set
$\Sigma = \Sigma_s \cup \Sigma_1$, 
with $\Sigma_s = \Sigma_f \cup \Sigma^d_s$, where $\K_s, \K_1$ are flat
and linear, $\K_s$ is a set of
$\Pi_0 \cup \Sigma_s$-clauses in which every variable occurs below a function
symbol in $\Sigma^d_s$, $\K_1$ a set of
$\Pi \cup \Sigma_1$-clauses in which every variable occurs below
an extension symbol in $\Sigma_1$.
Let $G$ be a set of flat and linear ground 
$\Pi^C$-clauses,  containing the symbols in $\Sigma_s$;  $C_s$ a
set of constants in $G$, 
and $\Gamma$ be the $\Pi_s^{\overline C_s}$-formula obtained after applying
Algorithm~1 to $\T$ and $G$, where $C_s \subseteq
{\overline C_s}$.
If all function symbols in $\Sigma^d_s$ are explicitly definable 
w.r.t.\ $\T$ over $\Pi_0 \cup \Sigma_f$, 
then we can compute 
the $\T$-general uniform interpolant for $G$ w.r.t.\ $\Sigma_s \cup
C_s$  using a reduction to computations of uniform interpolants 
in $\T_0 \cup {\sf UIF}_{\Sigma_f}$.  
\label{def-gen-un-int}
\end{thm}
{\em Proof:} 
If $\Gamma$ is the $\Pi_s^C$-formula obtained after applying
Algorithm~1 then, by Theorem~\ref{symb-elim-simplif}, 
$G \models_{\cal T} \Gamma$ and 
for every formula $\theta$ having only the symbols in 
$\Sigma_s \cup C_s \cup C$ in common  with $G$, 
if $G \models_{\T} \theta$ then $\Gamma \models_{\T}\theta$. 
Let $\Gamma_0$ be obtained from $\Gamma$ by replacing every term of
the form $f(c)$, where $f \in \Sigma_s$ and $c \in C$ with a new 
constant $c_{f(c)}$ and 
$\Gamma' = \Gamma_0 \wedge \bigwedge
c_{f(c)} = f(c)$, and 
$\Gamma'' = \Gamma_0 \wedge \bigwedge F_f(c, c_{f(c)})$, where $F_f$
is the explicit definition of $f$.
Then ${\cal T}_0 \cup {\sf UIF}_{\Sigma_f} \cup \K \models (\Gamma' \leftrightarrow \Gamma'')$.

\smallskip
\noindent Let ${\overline d}$ be all constants in $\Gamma'$ and $\Gamma''$ which are not in
$\Sigma_s \cup C_s$, and let $\psi$ be obtained by computing the $\T_0 \cup
{\sf UIF}_{\Sigma_f}$-uniform interpolant for $\Gamma''$ w.r.t.\
$C_s$. 
We prove that $\psi$ is a $\T$-general uniform interpolant for
$G$ w.r.t. $\Sigma_s \cup C_s$. 

\medskip
\noindent Clearly, $\psi$ is a
$\Pi_s^{C_s}$-formula. We prove that it satisfies the conditions of a
$\T$-general uniform interpolant of $G$. 

\medskip
\noindent (1) By Theorem~\ref{symb-elim-simplif} we know that $\Gamma$
is a $\T$-general uniform interpolant for $G$ w.r.t.\ $\Sigma_s$ and ${\overline
  C}_s$, so 
$G \models_{\T} \Gamma$. Let $\A$ be a
model of $\T$ and of $G$, and $\A^{C'}$ its expansion with the newly introduced 
symbols in ${\overline d}$ defined according to the definitions $c_{f(c)} =
f(c)$. Then $\A^{C'}$ is a model of $\T$ and of $\Gamma'$, hence 
it is a model of $\Gamma''$. Since $\psi$ is a $\T_0 \cup {\sf UIF}_{\Sigma_f}$-uniform interpolant of
$\Gamma''$ w.r.t.\ $C_s$, $\Gamma'' \models_{\T_0 \cup {\sf
    UIF}_{\Sigma_f}} \psi$, hence $\Gamma'' \models_{\T} \psi$. Thus, $\A$ is a model of $\psi$. 

\medskip
\noindent (2) We prove that for every $\T$-model $\A$ of $\psi$ there
exists a $\T$-model ${\cal D}$ of $G$ such that $\A_{|\Pi_r^{C_r}}$
embeds into ${\cal D}_{|\Pi_r^{C_r}}$, where $\Pi_r = \Pi_0 \cup
\Sigma_f \cup \Sigma_s$ and $C_r = C_s$. 

Let $\A$ be a model of $\T$ and of $\psi$. 
Since $\psi$ is the $\T_0 \cup
{\sf UIF}_{\Sigma_f}$-uniform interpolant of $\Gamma''$ w.r.t.\ $C_s$,
it is also a $\T_0 \cup {\sf UIF}_{\Sigma}$-uniform interpolant of $\Gamma''$ w.r.t.\ $C_s$.
Therefore, by Theorem~\ref{ghilardi-emb},
there exists another model $\B$ of $\T_0 \cup {\sf
  UIF}_{\Sigma}$ such that 
${\cal A}_{|\Pi}$ embeds into $\B_{|\Pi}$, the constants in $C_s$ have
the same interpretation in $\A$ and $\B$, and $\B \models \Gamma''$.

We construct a partial structure ${\cal P}$ 
starting from $\B$ as follows: The universe of ${\cal P}$ is $|\B|$. 
The symbols in
$\Pi_0 \cup \Sigma_f$ and the constants are defined as in $\B$. 
For every $f \in \Sigma \backslash \Sigma_f$, and all $v_1, \dots, v_n
\in|{\cal P}|$, 
$f_{{\cal P}}(v_1, \dots, v_n)$ is defined iff 
$v_1, \dots, v_n \in |\A|$, and in this case $f_{{\cal P}}(v_1, \dots, v_n) =
f_{\A}(v_1, \dots, v_n)$. 

Since $\A$ is a model of ${\cal T}_0 \cup {\sf  UIF}_{\Sigma_f} \cup
\K \cup \K_1$,  $\B$ is a model of $\T_0 \cup {\sf
  UIF}_{\Sigma}$, $\A_{\Pi}$ embeds into $\B_{\Pi}$, and $\K_s$ and
$\K_1$ do not contain common extension functions, we can prove 
that ${\cal P}$ is a weak partial model of ${\cal T}_0 \cup {\sf
  UIF}_{\Sigma_f} \cup \K \cup \K_1$ which is a total model of
$\T_0$. 
 By locality, ${\cal P}$ weakly embeds into a model 
${\cal C}$ of $\T_0 \cup {\sf  UIF}_{\Sigma_f} \cup \K \cup \K_1$. 

\medskip
\noindent Since $\Gamma''$ only contains symbols in $\Pi_0
\cup \Sigma_f \cup \overline{C_s}$,  
and their definitions in ${\cal C}$ are the same as
those in $\B$, ${\cal C}$ is a model of $\T_0 \cup {\sf
  UIF}_{\Sigma_f} \cup \K \cup \K_1$ which is also a model of 
$\Gamma''$. We know that $\T_0 \cup {\sf UIF}_{\Sigma_f} \cup \K_s
\cup \K_1 \models (\Gamma' \leftrightarrow \Gamma'')$, 
so $\C$ is also a model of $\Gamma'$. 
Since all the symbols in $\Gamma'$ are in $\Pi_0 \cup \Sigma_f \cup
\Sigma_s \cup C'$, their interpretation is the same in $\C$ as in
$\A^{C'}$, and $C \models \Gamma'$ it follows that $\C \models \Gamma$.

\medskip
\noindent By Theorem~\ref{gen-un-int-sem}, there exists another model, ${\cal
  D}$ of $\T_0 \cup {\sf
  UIF}_{\Sigma_1} \cup \K \cup \K_1$ which is a model of $G$, 
such that ${\cal C}_{|\Pi_r^{\overline{C_s}}}$ embeds into 
${\cal D}_{|\Pi_r^{\overline{C_s}}}$. 
By the way $\C$ is constructed, $\A_{|\Pi_r^{C_r}}$ embeds into ${\cal
  \C}_{|\Pi_r^{C_r}}$ hence into 
${\cal D}_{|\Pi_r^{C_r}}$.
 \QED

\medskip

\begin{ex} 
Consider the following chain of theory extensions of $\T_0 =
LI({\mathbb R})$:
$\T_0 \subseteq \T_0 \cup {\sf UIF}_h \subseteq \T_0 \cup {\sf UIF}_h
\cup \K$, where: 

\smallskip
$\K = \{ \forall x (x {\leq} 3
\rightarrow g(x) {\leq} f(x) {\leq} 3x), \forall x (x {>} 3 \rightarrow f(x)
\approx h(x)), \forall x (3x \approx g(x)) \}$. 

\smallskip
\noindent Let $\Sigma_s = \{ f, g, h \}$, $C_s = \{ u \}$. 
We compute $\exists c ( c \leq g(u) \wedge c \approx f(c) )$ as follows:  

\smallskip
\noindent {\bf We show that $f$ and $g$ are definable w.r.t.\ $\Pi_0
  \cup \{ h \}$.} 

\noindent We have the following chain of theory extensions:

\smallskip
$\T_0 \subseteq \T_0 \cup {\sf UIF}_{h} \subseteq \T_0 \cup {\sf
  UIF}_h \cup \K_g
\subseteq \T_0 \cup {\sf
  UIF}_h \cup \K_g \cup \K_f$ 

\smallskip
\noindent where $\K_f = \{ \forall x (x {\leq} 3
\rightarrow g(x) {\leq} f(x)  {\leq} 3x), \forall x (x {>} 3 \rightarrow f(x)
{\approx} h(x))\}$ and \\
$~~~~~~~$ $\K_g = \{ \forall x (3x {\approx} g(x)) \}$. 

\noindent All the extensions are local, $\K = \K_g \cup \K_f$,  
and the extension $\T_0 \subseteq \T_0 \cup \K$ is $\Psi$-local if we choose a
closure operator for computing instances as explained in the proof
of Theorem~\ref{thm:definability}.

\begin{itemize}
\item[1.] {\bf We show that $f$ is definable w.r.t.\ $\Pi_0 \cup \{ h
\}$.} \\
\begin{tabular}{@{}ll}
Let $\K' = \{$ & $\forall x (x \leq 3
\rightarrow g'(x) \leq f'(x)), \forall x (x \leq 3
\rightarrow f'(x) \leq 3x),$ \\
& $\forall x (x > 3 \rightarrow f'(x) \approx h(x)), \forall x (3x \approx
g'(x)) \}$ 
\end{tabular}

\smallskip
\noindent We have the following chain of theory extensions: 

\smallskip
$~~~~~~\T_0 \subseteq \T_0 \cup {\sf UIF}_h \subseteq \T_0 \cup {\sf UIF}_f
\cup \K \cup \K'$

\smallskip
This can be used for checking that $\T_0 \cup \K \cup \K' \models
\forall x (f(x) \approx f'(x))$ using the hierarchical reduction to satisfiability checking
w.r.t.\ $\T_0$ for (chains of) local theory extensions. 

\item[2.] {\bf We can prove that $g$ is definable w.r.t.\ $\Pi_0
\cup \{ h \}$}; we can compute the explicit definition: $F_g(a, b) := 3a \approx b$. 
\end{itemize}

\smallskip
\noindent {\bf We construct a explicit definition for $f$} by applying Algorithm~1: 

\smallskip
\begin{description}
\item[Step 1:] We instantiate and purify $\K \cup (f(a) \approx b)$: 

$\Psi({\sf est}(f(a) = b)) = \Psi(\{ f(a) \}) = \{ f(a), g(a), h(a) \}$

\smallskip
$\begin{array}{ll}
\K[\Psi({\sf est}(f(a) \approx b))] = \{ & (a \leq 3 \rightarrow g(a) \leq
f(a) \leq 3a), \\
&  (a > 3 \rightarrow f(a) \approx h(a)),  (3a \approx g(a)) \wedge f(a) \approx b \}.
\end{array}$

\smallskip
With the definition ${\sf Def} := \{ a_f \approx f(a), a_g \approx g(a), a_h \approx
h(a) \}$ we obtain after purification: 

\smallskip
$(a \leq 3 \rightarrow a_g \leq a_f \leq 3a) \wedge (a > 3 \rightarrow a_f \approx a_h) \wedge (3a \approx a_g) \wedge a_f \approx b$.

\smallskip
\item[Step 2:] We eliminate $a_f$, $a_g$, then replace $a_h$ back with $h(a)$ and obtain: 

$~~~F_f(a, b) := (a \leq 3 \rightarrow b \approx 3a) \wedge (a > 3
\rightarrow b \approx h(a))$.
\end{description}

\noindent {\bf To compute $\exists c ( c \leq g(u) \wedge c \approx
  f(c) )$} we replace $f(c)$ with $d_1$, $g(u)$ with
$d_2$ and obtain: 
$\exists d_1 \exists d_2 \exists c ( c \leq d_2 \wedge g(u) \approx
d_2 \wedge c \approx d_1 \wedge f(c) \approx d_1)$. 
We then replace the atoms $f(c) \approx d_1$ and $g(u) \approx d_2$ with the definitions and obtain: 

\smallskip
$\exists d_1 \exists d_2 \exists c ( c \leq d_2 \wedge 3u \approx d_2 \wedge
c \approx d_1 \wedge (c \leq 3
\rightarrow d_1 \approx 3c) \wedge (c > 3 \rightarrow d_1 \approx h(c)))$

$\begin{array}{ll}
\equiv & 0 \leq 3u \vee \exists d_1 (d_1 \leq 3u \wedge d_1 > 3 \wedge d_1 \approx h(d_1))
\end{array}$.

\smallskip
\noindent To eliminate $d_1$ we can 
use a method for uniform interpolation in $\T_0 \cup {\sf UIF}_{\{
  h \}}$.

\end{ex}

\section{Conclusions}
\label{conclusions}

We generalized the notion of cover \cite{Gulwani-Musuvathi} and uniform
  interpolant introduced in \cite{ghilardi-superposition} to a notion of general uniform
  interpolant, and proved a semantical characterization of uniform
  interpolants which generalizes a result in
  \cite{ghilardi-superposition}. 
We then proposed a method for symbol elimination in
local theory extensions and analyzed possibilities of using it 
for computing general uniform interpolants for ground formulae
w.r.t.\ such theories. 

Let $\T_0 \cup \K$ be a local theory extension with signature $\Pi$, 
a set $G$ of ground $\Pi^C$-clauses, a subset $\Sigma_s$ of extension
functions occurring in $G$ and a subset $C_s$ of constants occurring
in $G$.  
Our goal was to give a method for computing a 
$\T_0 \cup \K$-general uniform interpolant of $G$ w.r.t.\ $\Sigma_s
\cup C_s$ consisting of two steps: 
In a first step the function symbols not in $\Sigma_s$ were
eliminated, in a second step the constants not in $C_s$ were
eliminated (using methods for uniform interpolation). 

We identified situations in which the method for symbol elimination we
proposed could be used for eliminating the symbols not in $\Sigma_s$. 
After presenting a situation in which eliminating constants is
  problematic, we showed that general uniform
  interpolants can be computed in extensions with uninterpreted
  function symbols. 
We then studied two situations in which computing general uniform
interpolants can be ultimately reduced to computing uniform
interpolants in extensions of a theory with uninterpreted function symbols.

We first analyzed general uniform unification for ``semi-interpreted''
extension symbols. We defined a closure operator on shared symbols
(in which symbols occurring in the same clause in $\K$ with a shared 
symbols are also considered to be shared). 
We identified conditions (somewhat similar to the 
conditions on combinations of tame theories in
\cite{ghilardi-combinations}, or with the acyclicity conditions
considered in \cite{ghilardi-superposition}) under which the
computation of uniform interpolants can be reduced to the computation 
of uniform interpolants for extensions with uninterpreted function symbols. 
 
We then analyzed the situation in which some functions are 
definable w.r.t.\ the signature of the base theory together with 
uninterpreted function symbols, proposed a method for extracting 
explicit definitions for such functions which are implicitly
definable, 
and analyzed possibilities of computing uniform interpolants 
by eliminating definable functions (by replacing them with their
definitions), which makes it possible to use existing methods for 
uniform interpolation in extensions of theories with uninterpreted
function symbols.

Example~\ref{counterexample} shows that in some 
cases we cannot construct general uniform interpolants, but that 
some infinite sets of formulae might have the properties of general uniform  
interpolants. Such phenomena were studied in description 
logics, where it was shown that uniform interpolants do not always 
exist, but can often be expressed with the help of fixpoint operators. 
In \cite{ghilardi-superposition} a constraint superposition 
calculus was used for computing uniform interpolants in the 
theory of uninterpreted function symbols, and was proved to terminate.
However, constraint superposition calculi might not terminate in the 
presence of additional axioms.  Forms of symbol elimination based on using resolution
and/or superposition were used for interpolant generation in
\cite{hoder-kovacs-voronkov-10}, but it is not clear whether 
the method can be used for obtaining uniform interpolants.  
We hope that calculi such as the melting calculus \cite{HorbachW09}, 
which we studied in relationship with local 
theory extensions in \cite{horbach-sofronie13,horbach-sofronie14}, could be used 
for obtaining finite representations of infinite sets of clauses.

\medskip
\noindent {\bf Acknowledgment:} We thank the reviewers for their helpful comments. 
The research reported here was funded by the Deutsche
Forschungsgemeinschaft (DFG, German Research Foundation) – grant 465447331.

\bibliographystyle{abbrv}
\bibliography{sofronie-cade-2025-extended}

\begin{thebibliography}{10}

\bibitem{BruttomessoGR14}
R.~Bruttomesso, S.~Ghilardi, and S.~Ranise.
\newblock Quantifier-free interpolation in combinations of equality
  interpolating theories.
\newblock {\em {ACM} Trans. Comput. Log.}, 15(1):5:1--5:34, 2014.

\bibitem{ghilardi-superposition}
D.~Calvanese, S.~Ghilardi, A.~Gianola, M.~Montali, and A.~Rivkin.
\newblock Model completeness, uniform interpolants and superposition calculus.
\newblock {\em J. Autom. Reason.}, 65(7):941--969, 2021.

\bibitem{ghilardi-combinations}
D.~Calvanese, S.~Ghilardi, A.~Gianola, M.~Montali, and A.~Rivkin.
\newblock Combination of uniform interpolants via {Beth} definability.
\newblock {\em J. Autom. Reason.}, 66(3):409--435, 2022.

\bibitem{Ganzinger-01-lics}
H.~Ganzinger.
\newblock Relating semantic and proof-theoretic concepts for polynominal time
  decidability of uniform word problems.
\newblock In {\em 16th Annual {IEEE} Symposium on Logic in Computer Science,
  Boston, Massachusetts, USA, June 16-19, 2001, Proceedings}, pages 81--90.
  {IEEE} Computer Society, 2001.

\bibitem{ghilardi-dag}
S.~Ghilardi, A.~Gianola, and D.~Kapur.
\newblock Uniform interpolants in {EUF:} algorithms using dag-representations.
\newblock {\em Log. Methods Comput. Sci.}, 18(2), 2022.

\bibitem{Gulwani-Musuvathi}
S.~Gulwani and M.~Musuvathi.
\newblock Cover algorithms and their combination.
\newblock In S.~Drossopoulou, editor, {\em Programming Languages and Systems,
  17th European Symposium on Programming, {ESOP} 2008, Held as Part of the
  Joint European Conferences on Theory and Practice of Software, {ETAPS} 2008,
  Proceedings}, LNCS 4960, pages 193--207. Springer, 2008.

\bibitem{hoder-kovacs-voronkov-10}
K.~Hoder, L.~Kov{\'{a}}cs, and A.~Voronkov.
\newblock Interpolation and symbol elimination in {Vampire}.
\newblock In J.~Giesl and R.~H{\"{a}}hnle, editors, {\em Automated Reasoning,
  5th International Joint Conference, {IJCAR} 2010, Proceedings}, LNCS 6173,
  pages 188--195. Springer, 2010.

\bibitem{hodges}
W.~Hodges.
\newblock {\em Model theory}, volume~42 of {\em Encyclopedia of mathematics and
  its applications}.
\newblock Cambridge University Press, 1993.

\bibitem{horbach-sofronie13}
M.~Horbach and V.~Sofronie{-}Stokkermans.
\newblock Obtaining finite local theory axiomatizations via saturation.
\newblock In P.~Fontaine, C.~Ringeissen, and R.~A. Schmidt, editors, {\em
  Frontiers of Combining Systems - 9th International Symposium, FroCoS 2013,
  Proceedings}, LNCS 8152, pages 198--213. Springer, 2013.

\bibitem{horbach-sofronie14}
M.~Horbach and V.~Sofronie{-}Stokkermans.
\newblock Locality transfer: From constrained axiomatizations to reachability
  predicates.
\newblock In S.~Demri, D.~Kapur, and C.~Weidenbach, editors, {\em Automated
  Reasoning - 7th International Joint Conference, {IJCAR} 2014, Proceedings},
  LNCS 8562, pages 192--207. Springer, 2014.

\bibitem{HorbachW09}
M.~Horbach and C.~Weidenbach.
\newblock Deciding the inductive validity of {FOR} {ALL} {THERE} {EXISTS}
  \({}^{\mbox{*}}\) queries.
\newblock In E.~Gr{\"{a}}del and R.~Kahle, editors, {\em Computer Science
  Logic, 23rd international Workshop, {CSL} 2009, 18th Annual Conference of the
  EACSL, Proceedings}, LNCS 5771, pages 332--347. Springer, 2009.

\bibitem{ihlemann-jacobs-sofronie-tacas08}
C.~Ihlemann, S.~Jacobs, and V.~Sofronie{-}Stokkermans.
\newblock On local reasoning in verification.
\newblock In C.~R. Ramakrishnan and J.~Rehof, editors, {\em Tools and
  Algorithms for the Construction and Analysis of Systems, 14th International
  Conference, {TACAS} 2008, Held as Part of the Joint European Conferences on
  Theory and Practice of Software, {ETAPS} 2008, Proceedings}, LNCS 4963, pages
  265--281. Springer, 2008.

\bibitem{sofronie-ihlemann-10}
C.~Ihlemann and V.~Sofronie{-}Stokkermans.
\newblock On hierarchical reasoning in combinations of theories.
\newblock In J.~Giesl and R.~H{\"{a}}hnle, editors, {\em Automated Reasoning,
  5th International Joint Conference, {IJCAR} 2010, Proceedings}, LNCS 6173,
  pages 30--45. Springer, 2010.

\bibitem{McMillanRelationApproximation}
R.~Jhala and K.~L. McMillan.
\newblock Interpolant-based transition relation approximation.
\newblock In K.~Etessami and S.~K. Rajamani, editors, {\em Computer Aided
  Verification, 17th International Conference, {CAV} 2005, Proceedings}, LNCS
  3576, pages 39--51. Springer, 2005.

\bibitem{Kapur-et-all-06}
D.~Kapur, R.~Majumdar, and C.~G. Zarba.
\newblock Interpolation for data structures.
\newblock In M.~Young and P.~T. Devanbu, editors, {\em Proceedings of the 14th
  {ACM} {SIGSOFT} International Symposium on Foundations of Software
  Engineering, {FSE} 2006}, pages 105--116. {ACM}, 2006.

\bibitem{KonevWW09}
B.~Konev, D.~Walther, and F.~Wolter.
\newblock Forgetting and uniform interpolation in extensions of the description
  logic {${\cal E}{\cal L}$}.
\newblock In B.~C. Grau, I.~Horrocks, B.~Motik, and U.~Sattler, editors, {\em
  Proceedings of the 22nd International Workshop on Description Logics {(DL}
  2009)}, volume 477 of {\em {CEUR} Workshop Proceedings}. CEUR-WS.org, 2009.

\bibitem{konev-wolter-09}
B.~Konev, D.~Walther, and F.~Wolter.
\newblock Forgetting and uniform interpolation in large-scale description logic
  terminologies.
\newblock In C.~Boutilier, editor, {\em {IJCAI} 2009, Proceedings of the 21st
  International Joint Conference on Artificial Intelligence}, pages 830--835,
  2009.

\bibitem{lutz-wolter-ijcai11}
C.~Lutz and F.~Wolter.
\newblock Foundations for uniform interpolation and forgetting in expressive
  description logics.
\newblock In T.~Walsh, editor, {\em {IJCAI} 2011, Proceedings of the 22nd
  International Joint Conference on Artificial Intelligence}, pages 989--995.
  {IJCAI/AAAI}, 2011.

\bibitem{McAllester93}
D.~A. McAllester.
\newblock Automatic recognition of tractability in inference relations.
\newblock {\em J. {ACM}}, 40(2):284--303, 1993.

\bibitem{McMillanCAV03}
K.~L. McMillan.
\newblock Interpolation and {SAT}-based model checking.
\newblock In W.~{Hunt Jr}. and F.~Somenzi, editors, {\em Computer Aided
  Verification, 15th International Conference, {CAV} 2003, Proceedings}, LNCS
  2725, pages 1--13. Springer, 2003.

\bibitem{McMillanProver04}
K.~L. McMillan.
\newblock An interpolating theorem prover.
\newblock In K.~Jensen and A.~Podelski, editors, {\em Tools and Algorithms for
  the Construction and Analysis of Systems, 10th International Conference,
  {TACAS} 2004, Held as Part of the Joint European Conferences on Theory and
  Practice of Software, {ETAPS} 2004, Proceedings}, LNCS 2988, pages 16--30.
  Springer, 2004.

\bibitem{McMillanSurvey05}
K.~L. McMillan.
\newblock Applications of {Craig} interpolants in model checking.
\newblock In N.~Halbwachs and L.~D. Zuck, editors, {\em Tools and Algorithms
  for the Construction and Analysis of Systems, 11th International Conference,
  {TACAS} 2005, Held as Part of the Joint European Conferences on Theory and
  Practice of Software, {ETAPS} 2005, Proceedings}, LNCS 3440, pages 1--12.
  Springer, 2005.

\bibitem{peuter-sofronie19}
D.~Peuter and V.~Sofronie{-}Stokkermans.
\newblock On invariant synthesis for parametric systems.
\newblock In P.~Fontaine, editor, {\em Automated Deduction - {CADE} 27 - 27th
  International Conference on Automated Deduction, Proceedings}, LNCS 11716,
  pages 385--405. Springer, 2019.

\bibitem{peuter-sofronie-cade-2023}
D.~Peuter, V.~Sofronie{-}Stokkermans, and S.~Thunert.
\newblock On {$P$}-interpolation in local theory extensions and applications to
  the study of interpolation in the description logics {${\cal E}{\cal L}$,
  ${\cal E}{\cal L}^+$}.
\newblock In B.~Pientka and C.~Tinelli, editors, {\em Automated Deduction -
  {CADE} 29 - 29th International Conference on Automated Deduction,
  Proceedings}, LNCS 14132, pages 419--437. Springer, 2023.

\bibitem{sofronie-cade-05}
V.~Sofronie{-}Stokkermans.
\newblock Hierarchic reasoning in local theory extensions.
\newblock In R.~Nieuwenhuis, editor, {\em Automated Deduction - CADE-20, 20th
  International Conference on Automated Deduction, Proceedings}, LNCS 3632,
  pages 219--234. Springer, 2005.

\bibitem{Sofronie-ijcar-06}
V.~Sofronie{-}Stokkermans.
\newblock Interpolation in local theory extensions.
\newblock In U.~Furbach and N.~Shankar, editors, {\em Automated Reasoning,
  Third International Joint Conference, IJCAR 2006, Proceedings}, LNCS 4130,
  pages 235--250. Springer, 2006.

\bibitem{sofronie-lmcs08}
V.~Sofronie{-}Stokkermans.
\newblock Interpolation in local theory extensions.
\newblock {\em Logical Methods in Computer Science}, 4(4), 2008.

\bibitem{sofronie-ijcar10}
V.~Sofronie{-}Stokkermans.
\newblock Hierarchical reasoning for the verification of parametric systems.
\newblock In J.~Giesl and R.~H{\"{a}}hnle, editors, {\em Automated Reasoning,
  5th International Joint Conference, {IJCAR} 2010, Proceedings}, LNCS 6173,
  pages 171--187. Springer, 2010.

\bibitem{sofronie-cade13}
V.~Sofronie{-}Stokkermans.
\newblock Hierarchical reasoning and model generation for the verification of
  parametric hybrid systems.
\newblock In M.~P. Bonacina, editor, {\em Automated Deduction - {CADE-24} -
  24th International Conference on Automated Deduction, Proceedings}, LNCS
  7898, pages 360--376. Springer, 2013.

\bibitem{sofronie-ijcar16}
V.~Sofronie{-}Stokkermans.
\newblock On interpolation and symbol elimination in theory extensions.
\newblock In N.~Olivetti and A.~Tiwari, editors, {\em Automated Reasoning - 8th
  International Joint Conference, {IJCAR} 2016, Proceedings}, LNCS 9706, pages
  273--289. Springer, 2016.

\bibitem{sofronie-fuin-2017}
V.~Sofronie{-}Stokkermans.
\newblock Representation theorems and locality for subsumption testing and
  interpolation in the description logics $\cal{EL}$, $\cal{EL}^+$ and their
  extensions with \emph{n}-ary roles and numerical domains.
\newblock {\em Fundamenta Informaticae}, 156(3-4):361--411, 2017.

\bibitem{sofronie-lmcs18}
V.~Sofronie{-}Stokkermans.
\newblock On interpolation and symbol elimination in theory extensions.
\newblock {\em Log. Methods Comput. Sci.}, 14(3), 2018.

\bibitem{sofronie-fundamenta20}
V.~Sofronie{-}Stokkermans.
\newblock Parametric systems: Verification and synthesis.
\newblock {\em Fundamenta Informaticae}, 173(2-3):91--138, 2020.

\bibitem{sofronie-cade-2025}
V.~Sofronie-Stokkermans.
\newblock On symbol elimination and uniform interpolation in theory extensions.
\newblock In {\em Automated Deduction - {CADE} 30 - 30th International
  Conference on Automated Deduction, Proceedings}. Springer, 2025.
\newblock To appear.

\bibitem{Sofronie-Ihlemann07}
V.~Sofronie{-}Stokkermans and C.~Ihlemann.
\newblock Automated reasoning in some local extensions of ordered structures.
\newblock In {\em 37th International Symposium on Multiple-Valued Logic,
  {ISMVL} 2007, 13-16 May 2007, Oslo, Norway}. {IEEE} Computer Society, 2007.

\bibitem{sofronie-ihlemann-ismvl-07}
V.~Sofronie{-}Stokkermans and C.~Ihlemann.
\newblock Automated reasoning in some local extensions of ordered structures.
\newblock {\em Multiple-Valued Logic and Soft Computing}, 13(4-6):397--414,
  2007.

\end{thebibliography}

\newpage
\appendix
\vspace{-2mm}
\section{Local theory extensions}
\label{app:local}
 
\vspace{-2mm}
In \cite{sofronie-cade-05} we proved that if 
every weak partial model of an extension 
${\mathcal T}_0 \cup {\mathcal K}$ of a base theory ${\mathcal T}_0$ 
with total base functions can be embedded into a total
model of the extension, then the extension is local. 
In \cite{ihlemann-jacobs-sofronie-tacas08} we lifted these results to
$\Psi$-locality. 
In this appendix we give a short summary of the main definitions and 
results which can be useful for understanding some of the proofs in
this paper. 

\smallskip
\noindent {\bf Partial structures.}
\noindent In \cite{sofronie-cade-05} we showed that 
local theory extensions can be 
recognized by showing that certain partial models embed into total
ones, and in \cite{sofronie-ihlemann-10} we established similar
results for $\Psi$-local theory extensions and generalizations
thereof. 
We introduce the main  
definitions here.

Let $\Pi = (\Sigma, {\sf Pred})$ be a
first-order signature with set of function symbols $\Sigma$ 
and set of predicate symbols ${\sf Pred}$. 
A \emph{partial $\Pi$-structure} is a structure 
$\A = (A, \{f_\A\}_{f\in\Sigma}, \{P_\A\}_{P\in \Pred})$, 
where $A$ is a non-empty set, for every $n$-ary $f \in \Sigma$, 
$f_\A$ is a partial function from $A^n$ to $A$, and for every $n$-ary 
$P \in {\sf Pred}$, $P_\A \subseteq A^n$. We consider constants (0-ary functions) to be always 
defined. $\A$ is called a \emph{total structure} if the 
functions $f_\A$ are all total. 
Given a (total or partial) $\Pi$-structure $\A$ and $\Pi_0 \subseteq \Pi$ 
we denote the reduct of 
$\A$ to $\Pi_0$ by $\A{|_{\Pi_0}}$.

The notion of evaluating a term $t$ with variables $X$ w.r.t.\ 
an assignment $\beta : X \rightarrow A$
 for its variables in a partial structure
$\A$ is the same as for total algebras, except that the evaluation is
undefined if $t = f(t_1,\ldots,t_n)$ 
and at least one of
$\beta(t_i)$ is undefined, or else $(\beta(t_1),\ldots,\beta(t_n))$ is
not in the domain of $f_\A$.

\vspace{-1mm}
\begin{defi}
A \emph{weak $\Pi$-embedding} between two partial $\Pi$-structures
$\A$ and $\B$, where $\A = (A, \{f_\A\}_{f\in \Sigma}, \{P_\A\}_{P \in \Pred})$
 and  $\B = (B, \{f_\B\}_{f\in \Sigma}, \{P_\B\}_{P \in \Pred})$
is a total map $\varphi : A \rightarrow B$ such that 
\begin{enumerate}
\vspace{-2mm}
\item[(i)] $\varphi$ is
an embedding w.r.t.\ ${\sf Pred} \cup \{ = \}$, i.e.\ 
 for every $P \in {\sf Pred}$ with arity $n$ and every 
$a_1, \dots, a_n \in \A$, 
$(a_1, \dots, a_n) \in P_\A$ iff $(\varphi(a_1), \dots, \varphi(a_n))\in P_\B$. 
\item[(ii)]  
whenever $f_\A(a_1, \dots, a_n)$ is defined (in $\A$),  then 
$f_\B(\varphi(a_1), \dots, \varphi(a_n))$ is defined (in $\B$) and 
$\varphi(f_\A(a_1, \dots, a_n)) = f_\B(\varphi(a_1), \dots, \varphi(a_n))$,
for all $f \in \Sigma$.
\vspace{-2mm} 
\end{enumerate}
\end{defi} 
%

\vspace{-2mm}
\begin{defi}[Weak validity]
Let $\A$ be a partial $\Pi$-algebra
and $\beta : X {\rightarrow} A$ a valuation for its variables.
$(\A, \beta)$ {\em weakly satisfies a clause  $C$} (notation: 
$(\A, \beta) \models_w C$) if either some of the literals in 
$\beta(C)$ are not defined or otherwise all literals are defined and
for at least one literal $L$ in $C$, $L$ is true in $\A$ w.r.t.\ $\beta$. 
$\A$ is a {\em weak partial model} 
of a set of clauses ${\mathcal K}$ if $(\A, \beta)  \models_w C$ for every 
valuation 
$\beta$ and every clause $C$ in ${\mathcal K}$. 
\end{defi}

\noindent {\bf Recognizing $\Psi$-theory extensions.}
%
Let ${\cal A} = (A, \{ f_{\cal A} \}_{f \in \Sigma_0 \cup \Sigma} \cup
C, \{ P_{\cal A}
\}_{P \in {\sf Pred}})$ be a partial $\Pi^C$-structure with total
$\Sigma_0$-functions. 
Let $\Pi^A$ be the extension of the signature $\Pi$ with constants 
from $A$. We denote by $T({\cal A})$ the following set of ground 
$\Pi^A$-terms:
 
\smallskip
$T(\A) := \{ f(a_1,...,a_n) \,|\; f \in \Sigma, a_i \in A, i=1,\dots,n, f_{\A}(a_1,...,a_n) \text{ is defined }  \}. $

\smallskip
\noindent Let ${\sf PMod}_{w,f}^\Psi({\Sigma}, {\mathcal T})$ be the class of all
weak partial models $\A$ of ${\mathcal T}_0 \cup {\mathcal K}$, such that
$\A{|_{\Pi_0}}$ is a total model of $\T_0$, the
$\Sigma$-functions are possibly partial, $T(\A)$ is finite and 
all terms in $\Psi({\sf est}(\K, T(\A)))$ are
defined (in the extension $\A^A$ with
constants from $A$).
We consider the following embeddability property:

\vspace{-1mm}
\begin{tabbing}
\= $({\sf Emb}_{w,f}^\Psi)$ \quad \= Every ${\cal A} \in {\sf PMod}_{w,f}^\Psi({\Sigma},  \T)$ weakly embeds into a total model of $\T$.
\index{$\Psi$-!embeddability}
%
%
\end{tabbing}

\ignore{\medskip
\noindent 
We also consider the properties $({\sf EEmb}_{w,f}^{\Psi})$,  
which additionally requires the embedding to be {\em elementary} and 
$({\sf Comp}_f)$  which requires that every  structure $\alg{A} \in {\sf
  PMod}_{w,f}^\Psi({\Sigma}, \T)$ embeds
into a total model of $\T$ {\em with the same support}. 
} 

\noindent When establishing links between locality and embeddability we require 
that the clauses in $\K$
are \emph{flat} 
and \emph{linear} w.r.t.\ $\Sigma$-functions.
When defining these notions we distinguish between ground and 
non-ground clauses.

\vspace{-1mm}
\begin{defi}
An {\em extension clause $D$ is flat} (resp. \emph{quasi-flat}) 
when all symbols 
below a $\Sigma$-function symbol in $D$ are variables 
(resp. variables or ground $\Pi_0$-terms).
$D$ is \emph{linear}  if whenever a variable occurs in two terms of $D$
starting with $\Sigma$-functions, the terms are equal, and 
no term contains two occurrences of a variable.

A {\em ground clause $D$ is flat} if all symbols below a $\Sigma$-function 
in $D$ are constants.
A {\em ground clause $D$ is linear} if whenever a constant occurs in two terms in $D$ whose root symbol is in $\Sigma$, the two terms are identical, and
if no term which starts with a $\Sigma$-function contains two occurrences of the same constant.
\label{flat}
\vspace{-1mm}
\end{defi} 
 \begin{defi}[\cite{sofronie-ihlemann-10}]
\vspace{-1mm}
With the above notations, let $\Psi$ be a map associating with 
$\K$ and a set of $\Pi^C$-ground terms $T$ 
a set $\pK(T)$ of $\Pi^C$-ground terms. 
We call $\pK$ a \emph{term closure operator} if for all sets of ground
terms $T, T'$ we have: 
\begin{enumerate}
\item $\mathrm{est}(\K, T) \subseteq \pK(T)$,
\item $T \subseteq T' \Rightarrow \pK(T) \subseteq \pK(T')$,
\item $\pK(\pK(T)) \subseteq \pK(T)$,
\item for
  any map $h: C \rightarrow C$, $\bar{h}(\pK(T)) = \Psi_{\bar{h}\K}(\bar{h}(T))$,
  where $\bar{h}$ is the canonical extension of $h$ to extension
  ground terms.
\end{enumerate}
\vspace{-1mm}
\end{defi}

\begin{thm}[\cite{ihlemann-jacobs-sofronie-tacas08,sofronie-ihlemann-10}]
Let ${\mathcal T}_0$ be a first-order theory and $\K$ a set of universally closed flat clauses in the signature
$\Pi$. The following hold: 
\vspace{-1mm}
\begin{enumerate}
\item If all clauses in $\K$ are linear  
and $\Psi$ is a term closure operator with the property 
that for every flat set of ground terms $T$, $\Psi(T)$ is
flat then 
condition $({\sf Emb}_{w,f}^\Psi)$ 
implies $({\sf Loc}_f^{\Psi})$.
\item If  the extension ${\mathcal T}_0 \subseteq {\mathcal T} {=} {\mathcal T}_0 {\cup} {\mathcal K}$
satisfies $({\sf Loc}_f^{\Psi})$ then $({\sf Emb}_{w,f}^\Psi)$ holds. 
\end{enumerate}
\vspace{-1mm}
\label{check-loc}
\end{thm}
We now prove a result which was at times used in this paper: 
\begin{lemma}
Let $\T_0 \subseteq \T_0 \cup \K_1 \subseteq \T_0 \cup \K_1 \cup \K_2$
be a chain of local theory extensions, such that $\K_1, \K_2$ are flat
and linear. 
Then $\T_0 \cup \K_1 \subseteq \T_0 \cup \K_1 \cup \K_2$ is a local
theory extension. 
\end{lemma}
{\em Proof:} Let ${\cal P}$ be a weak partial model of $\T_0 \cup \K_1
\cup \K_2$ which is a total model of $\T_0 \cup \K_1$. 
Then ${\cal T}$ is a weak partial model of $\T_0 \cup \K_1
\cup \K_2$ which is a total model of $\T_0$, hence it embeds into a
total model of $\T_0 \cup \K_1
\cup \K_2$. This proves that $\T_0 \cup \K_1 \subseteq \T_0 \cup \K_1 \cup \K_2$ is a local
theory extension. \QED

\bigskip
\noindent {\bf Chains of theory
  extensions, hierarchical reasoning, closure operators.}
Let $\T_0$ be a  theory with signature $\Pi_0 = (\Sigma_0, {\sf
  Pred})$, and let 
$\Sigma_1, \Sigma_2$ be disjoint sets of additional function symbols. 
Let $\T_0 \subseteq \T_0 \cup \K_1 \subseteq \T_0 \cup \K_1 \cup \K_2$
be a chain of local theory extensions, such that $\K_1, \K_2$ are sets
of flat
and linear clauses,  $\K_1$ is a set of $\Pi_0 \cup \Sigma_1$-clauses
with the property that every variable occurs below a function symbol
in $\Sigma_1$, and $\K_2$ is a set of  $\Pi_0 \cup \Sigma_1 \cup \Sigma_2$-clauses
with the property that every variable occurs below a function symbol
in $\Sigma_2$. 

\smallskip
\noindent Let $G$ be a set of flat and linear $(\Pi_0 \cup \Sigma_1 \cup
\Sigma_2)^C$-clauses. We can check whether $\T_0 \cup \K_1 \cup \K_2 \cup G \models \perp$
as follows: 

\smallskip
\noindent By the locality of the extension $\T_0 \cup \K_1 \subseteq \T_0
  \cup \K_1 \cup \K_2$, the following are equivalent: 
\begin{itemize}
\item[(a)] $\T_0 \cup \K_1 \cup \K_2 \cup G \models
  \perp$. 
\item[(b)] $\T_0 \cup \K_1 \cup \K_2[G] \cup G \models \perp$.
\item[(c)] $\T_0 \cup \K_1 \cup {\K_2}_0 \cup G_0 \cup {\sf Con}^{\Sigma_2}_0
  \models \perp$, where ${\K_2}_0 \cup G_0 \cup {\sf Con}^{\Sigma_2}_0$ is
  obtained from $\K_2[G] \cup G$ as indicated by Theorem~\ref{lemma-rel-transl}. 
(Since every variable of a clause in $\K_2$ occurs below a function
  symbol in $\Sigma_2$, $\K_2[G]$ is a set of ground clauses, 
the $\Sigma_2$-terms are replaced by fresh constants in $C$ and ${\sf
  Con}^{\Sigma_2}_0$
are the corresponding congruence axioms for these $\Sigma_2$-terms.)
\end{itemize}
\noindent Let $G^1 := {\K_2}_0 \cup G_0 \cup {\sf Con}^{\Sigma_2}_0$. Then $G^1$ is
  a set of ground $(\Pi \cup \Sigma_1)^C$-clauses, and by the locality of the extension
$\T_0 \subseteq \T_0 \cup \K_1$ the following are equivalent: 
\begin{itemize}
\item[(c)] $\T_0 \cup \K_1 \cup G^1 \models \perp$.
\item[(d)] $\T_0 \cup \K_1[G^1] \cup G^1 \models \perp$.
\item[(e)] $\T_0 \cup {\K_1}_0 \cup G^1_0 \cup {\sf Con}^{\Sigma_1}_0
  \models \perp$
where ${\K_1}_0 \cup G^1_0 \cup {\sf Con}^{\Sigma_1}_0$ is
  obtained from $\K_1[G^1] \cup G^1$ as indicated by Theorem~\ref{lemma-rel-transl}. 
(Since every variable of a clause in $\K_1$ occurs below a function
  symbol in $\Sigma_1$, $\K_1[G^1]$ is a set of ground clauses, 
the $\Sigma_1$-terms are replaced by fresh constants in $C$ and ${\sf
  Con}^{\Sigma_1}_0$
are the corresponding congruence axioms for these $\Sigma_1$-terms.)
\end{itemize}
The terms needed in the
instantiation can be described as follows:

\noindent Let $T_2(G) {=} {\sf est}_{\Sigma_2}(\K_2, G)$ be the set of ground
  $\Sigma_2$-terms occurring in $\K_2$ and $G$, 
and $T_1(G)$ the set of all ground  $\Sigma_1$-terms occuring
  in $\K_2[G] {=} \K_2[T_2(G)]$ and $G$.

\smallskip
\noindent The set of all instances needed for checking the satisfiability
  of $G$ with respect to  $\T_0 \cup \K_1 \cup \K_2$ is $\Psi(G) = T_2(G) \cup
  T_1(G)$.

\begin{ex}
Consider the following chain of theory extensions:

\smallskip
$\T_0 \subseteq \T_0 \cup \K_1
\subseteq \T_0 \cup \K_1 \cup \K_2$ 

\smallskip
\noindent where $\K_2 = \{ \forall x (x \leq 3
\rightarrow g(x) \approx f(x)), \forall x (x > 3 \rightarrow f(x)
\approx x)\}$,  \\
$~~~~~~~$  $\K_1 = \{ \forall x (3x \approx g(x)) \}$,\\
and $\T_0 = LI({\mathbb R})$, and let $G = f(c) \not\approx 3c \wedge f(c) \not\approx c$. 

\noindent Then $T_2(G) = {\sf est}_{\{f \}}(\K_2, G) = \{ f(c) \}$, and $\K_2[G] = \{ c \leq 3
\rightarrow g(c) \approx f(c), \\
c > 3 \rightarrow f(c) \approx c \}$.
Since ${\sf est}_{\{g \}}(\K_2[G]) = \{ g(c) \}$,  we have
$T_1(G) := \{ g(c) \}$. \\
The set of ground terms needed for checking the
satisfiability of $G$ w.r.t.\ \\
$\T_0 \cup \K_1 \cup \K_2$ is 
$\Psi(G) = T_2(G) \cup T_1(G) = \{ f(c), g(c)\}$. 
\end{ex}

\end{document}